\documentclass[12pt]{thesis}

\usepackage{times}

\usepackage{graphicx}
\usepackage{rotating}
\usepackage{makeidx}
\usepackage{subfig}
\usepackage{wrapfig}

\usepackage{cite}
\usepackage{algorithmic}
\usepackage{graphicx}
\usepackage{xcolor}
\usepackage{amsmath,amsfonts,amssymb}
\newcommand{\mc}{\mathcal}
\newcommand{\norm}[1]{\left\lVert#1\right\rVert}
\usepackage{float}

\doctype{thesis}
\title{UAV Control Optimization via Decentralized Markov Decision Processes}
\author{Md Ali Azam}
\degree{Master of Science in Electrical Engineering}
\defensedate{November 16, 2020}
\gradyear{2020}
\department{Electrical Engineering}

\signatureline{Major Professor --- Shankarachary Ragi, Ph.D., Department of Electrical Engineering}

\signatureline{Graduate Division Representative --- Randy C Hoover, Ph.D.,
  Department of Computer Science and Engineering}

\signatureline{Committee Member --- Sayan Roy, Ph.D., Department of
  Electrical Engineering}

\signatureline{Head of the Electrical Engineering Department --- Thomas P Montoya, Ph.D.}

\signatureline{Dean of Graduate Education --- Maribeth H Price, Ph.D.}

\begin{document}

\begin{figure}
\centering{\includegraphics[width= \columnwidth, trim = 102 10 10 10,clip]{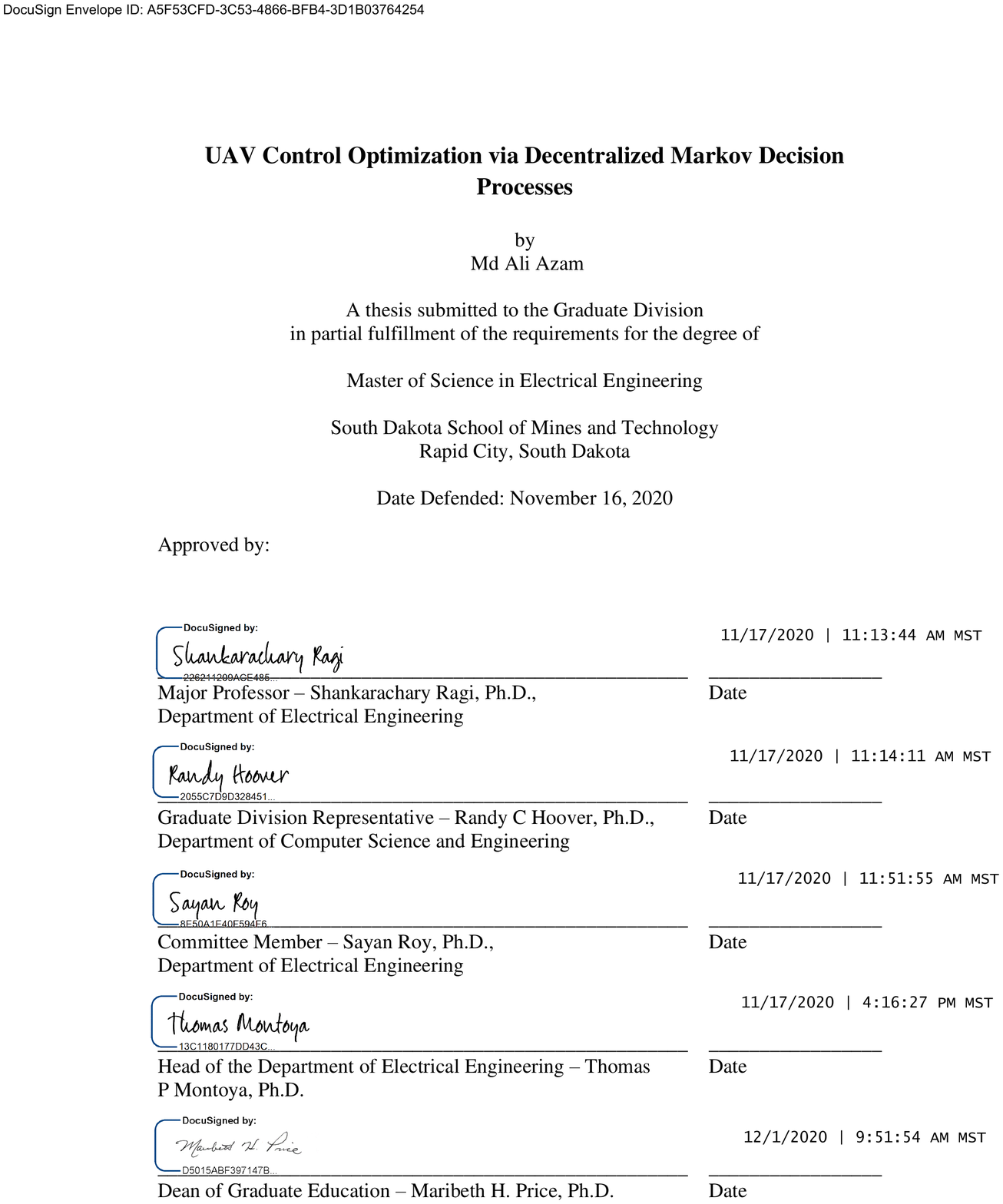}}
\end{figure}


\makecopyright 

\preliminaries

\begin{abstract}

Unmanned aerial vehicle (UAV) swarm control has applications including target tracking, surveillance, terrain mapping, and precision agriculture. Decentralized control methods are particularly useful when the swarm is large, as centralized methods (a single command center/computer controlling the UAVs) suffer from exponential computational complexity, i.e., the computing time to obtain the optimal control for the UAVs grow exponentially with the number of UAVs in the swarm in centralized approaches. Although many centralized control methods exist, literature lacks decentralized control frameworks with broad applicability. To address this knowledge gap, we present a novel decentralized UAV swarm control strategy using a decision-theoretic framework called \emph{decentralized Markov decision process} (Dec-MDP). We build these control strategies in the context of two case studies: a) swarm formation control problem; b) swarm control for multitarget tracking. As most decision theoretic formulations suffer from the curse of dimensionality, we adapt an approximate dynamic programming method called \emph{nominal belief-state optimization} (NBO) to solve the decentralized control problems approximately in both the case studies. In the formation control case study, the objective is to drive the swarm from a geographical region to another geographical region where the swarm must form a certain geometrical shape (e.g., selected location on the surface of a sphere). The motivation for studying such problems comes from data fusion applications with UAV swarms where the fusion performance depends on the strategic relative separation of the UAVs from each other. In the target tracking case study, the objective is the control the motion of the UAVs in a decentralized manner while maximizing the overall target tracking performance. Motivation for this case study comes from the surveillance applications using UAV swarms. 

Furthermore, we develop an average consensus-based decentralized data fusion approach for using data generated by the UAVs in the context of target tracking applications. In centralized control setting, often we use Bayesian-based data fusion strategies. There is no consensus in the literature on what data fusion strategy works best in decentralized control problems. To address this knowledge gap to an extent, we extend the \textit{average consensus} algorithm to fuse the local state estimate information with that of the neighbors when the UAVs pass information for data fusion while optimizing the controls in the decentralized setting. We test the performance of our consensus-based data fusion approach in various UAV swarm network configurations and assess its performance against standard Bayesian data fusion approaches. 

\end{abstract}

\begin{acknowledgments}
Firstly, I would like to express my sincere gratitude to my advisor Dr. Shankarachary Ragi for his continuous support and motivation during my MS study and related research. The door of Dr. Ragi was always open whenever I needed help with my research. I remember when I started writing my first paper, I barely had context in my paper draft. Dr. Ragi showed me how to write technical content line by line during my entire MS study. I will always be grateful to him for his incessant support and directions. He steered me in the right direction whenever he thought I needed it. I could not have imagined having a better supervisor for MS study.   

Besides my advisor, I would like to thank the rest of my thesis committee: Dr. Randy C Hoover, and Dr. Sayan Roy for their insightful comments and encouragement. My sincere thanks to Dr. Hoover for occasional discussion on any research and course related problems. I have learned so many things from Dr. Hoover for which I am truly indebted to him. 

I thank my labmates and fellow researchers for the stimulating discussions on research problems, motivations, and support throughout my MS study. I also sincerely convey my heartfelt thanks and gratitude to all current and past Bangladeshi students at SDSMT for their loves and supports along the way.

Last but not the least, I would like to thank my family members for supporting me from the beginning of my journey. Their inspirations and loves have brought me here today. 

\end{acknowledgments}

\tableofcontents

\listoftables

\listoffigures





\body

\chapter{Introduction}

Unmanned Aerial Vehicle (UAV) swarm formation has applications in many areas of research such as infrastructure inspection \cite{detection} and surveillance \cite{surveillance}, target tracking \cite{tracking}, and precision agriculture. UAV swarm formation and information passing for data fusion within the swarm requires control and optimization strategies that can be implemented in near real-time. Many centralized control \cite{zhan2005, hao2016, zhao, ragi-path-planning, chong2009} strategies exist, albeit suffer from exponential computational complexities. To address this challenge, we develop decentralized control and study data fusion methods using a decision theoretic framework called \emph{decentralized Markov decision process} (Dec-MDP). We will develop these methods in two case studies: a) decentralized formation control \cite{zhao,viana,tracking} of a UAV swarm; b) UAV motion control for multitarget tracking \cite{LaValle2006,ragi-tracking,Fortmann1980, Andri2012, Vermaak2005, Benfold2011}. In the first case study, the primary objective is to drive the swarm fly and hover in a certain geographical region while forming a certain geometrical shape. The formation shape of a swarm can be useful in many ways. The motivation for studying such problems comes from data fusion applications with UAV swarms where the fusion performance depends on the strategic relative separation of the UAVs from each other. In the target tracking case study, the objective is the control the motion of the UAVs in a decentralized manner while maximizing the overall target tracking performance. Motivation for this case study comes from the surveillance applications of UAV swarms. We develop our control and data fusion strategies in two-dimensional (2D) simulation scenarios in this study, which can be easily extended to 3D.

Formation control is one of the most actively studied topics in multi-agent systems and swarm intelligence. Different formation control settings have been studied in the past: ground vehicles \cite{Das2002, Fax2004, Ghabcheloo2006}, unmanned aerial vehicles (UAVs) \cite{Singh2000, Koo2001}, surface and underwater autonomous vehicles (AUVs) \cite{Edwards2004, Skjetne2002}. Regardless of settings, there are many different methodologies developed by the researchers to tackle formation control problem, e.g.,  behavior-based, virtual structure, and leader following. The authors of \cite{Balch1998, Lawton2003} developed behavior-based approach where they described desired behavior for each robot, e.g., collision avoidance, formation keeping, and target seeking. The control commands for the robot is determined by weighing the relative importance of each behavior. The virtual structure approach \cite{Do2007, Lewis1997} takes a physical object shape as a reference and mimics the formation of that shape. The robots are required to communicate with each other in order to achieve a formation in this approach which requires significant communication costs (e.g., bandwidth). The leader following approach \cite{Das2002} requires a robot, assigned as a leader, moves according to a predetermined trajectory. The other robots, the followers, are designed to follow the leader maintaining a desired distance and orientation with respect to the leader. The main drawback of this approach is that the followers are dependent on the leader to achieve the goal (formation). The system may collapse if the leader fails when possibly the leader runs short on power or when the communications link fails. Considering the aforementioned limitations of formation control, which specifically stem from centralized approaches, we develop a decentralized Markov decision process (Dec-MDP) based formation control approach for a UAV swarm. Our decentralized control strategies are robust to failures of individual UAVs in the swarm and also robust to communications link failures.  

Centralized control strategies for UAV swarm control are well studied \cite{zhan2005, hao2016, zhao, ragi-dec-pomdp, chong2009}. For instance, the authors of \cite{zhan2005,ragi-tracking} developed UAV control strategies for target tracking in a centralized setting. In centralized systems like these, typically, there exists a notional fusion center (a computing node) that collects and fuses the sensor measurements (e.g., using Bayes' theorem) from all the UAVs and runs a tracking algorithm (e.g., Kalman filter) to maintain and update the estimate of the state of the system. More importantly, the fusion center computes the combined optimal control commands for all the UAVs to maximize the system performance. For instance, the authors of \cite{ragi-path-planning} used the notion of fusion center to control fixed-wing UAVs for multitarget tracking while accounting for collision avoidance and wind disturbance on UAVs. Although, these centralized control and fusion strategies are easy to implement, they are computationally expensive especially if the swarm is large. Specifically, the computational complexity increases exponentially with the number of UAVs in the swarm.  

To tackle these challenges, a few studies in the literature developed decentralized control strategies \cite{ragi-dec-pomdp, kim2016, wei2014, Bakule2008}. The authors of \cite{ragi-dec-pomdp} used the decentralized partially observable Markov decision process (Dec-POMDP) to formulate and solve a target tracking problem with a swarm of decentralized UAVs. As solving decentralized POMDP is very hard (as is the case with solving any decision-theoretic methods), the authors introduced an approximate dynamic programming method called \emph{nominal belief-state optimization} (NBO) to solve the control problem. The authors in \cite{viana} developed a UAV formation control approach using decentralized Model Predictive Control (MPC). In their work, the UAVs were able to avoid collisions with multiple obstacles in a decentralized manner. They used figure of eight as reference trajectory; their results show that the UAVs were able to avoid collision with obstacles and among themselves. Several recent papers describe formation control of different geometric shapes, e.g., multi-agent circular shape with a leader \cite{zhao}. The authors of \cite{zhao} propose centralized formation control, which is not suitable for swarm control when the number of UAVs in the swarm is large. Although decentralized control methods exist in the literature, our method is novel in the sense that each UAV in the swarm optimizes its own control commands and its nearest neighbor's controls over time. Then, each UAV implements its own optimized controls, and discards the neighbor's controls. We anticipate, from this decentralized control optimization approach, a global cooperative behavior among the UAVs emerges mimicking a centralized control approach. The authors of \cite{Pham2017} demonstrated a successful use of a distributed UAV control framework for wildfire monitoring while avoiding in-flight collisions. The authors of \cite{Zhang2013} introduced path tracking and desired formation for networked mobile vehicles using non-linear control theory to maintain the formation in the network. They have showed that path tracking error of each vehicle is reduced to zero and formation is achieved asymptotically. As centralized control strategies suffer from exponential computational complexity and high memory usage, the decentralized control methods are being actively pursued in the context of swarm control, especially when the size of the swarm is large. A survey of these decentralized control strategies can be found in \cite{Bakule2008}. With this motivation, we adapt a decision theoretic framework called decentralized Markov decision process (Dec-MDP) to solve our UAV swarm control problems. As Dec-MDPs suffer from the curse of dimensionality, we extend a fast heuristic approach called \emph{nominal belief-state optimization} (NBO) \cite{miller2009, ragi-path-planning} to solve the Dec-MDP problem to obtain suboptimal but near real-time solutions for UAV swarm control. 

In decentralized swarm systems, information fusion is as important as the control strategies. Since there is no central fusion center in decentralized systems, there is a need for methods that allows the agents/sensors to cooperatively share information among themselves and perform information fusion locally. To this end, we evaluate the performance of an \textit{average consensus algorithm} for multisensor fusion for a target tracking application. Multisensor data fusion has been widely studied \cite{Hall1997, Khaleghi2013, Sun2004} in the context of applications including surveillance, remote sensing, and guidance and control of autonomous vehicles. A commonly used stochastic signal processing method Kalman filter \cite{Kalman1960} allows multisensor data fusion for linear systems, where the process and measurement noise are modeled by zero-mean Gaussian distributions. The Kalman filter based multisensor fusion requires prior knowledge of the cross covariance of estimate errors. In this study, we will perform a comparative study to compare the performance of consensus-based information fusion against Kalman filter-based data fusion for a dencentralized network of sensors in a target tracking context.  

The authors of \cite{Shen2010, Matei2012} developed distributed fusion algorithms to estimate the state of interest effectively. In distributed fusion approach, each sensor has its own local information and combine with other sensors of the same network to update its local information. The authors of \cite{Battistelli2013} developed a novel consensus approach called Gaussian Mixture-Cardinalized Probability Hypothesis Density (GM-CPHD) filter for multitarget tracking application. Despite many distributed fusion approaches, some consensus approaches \cite{Olfati2004, Olfati2007, Ren2007, nedic2010, nedich} are very successful in homogeneous data fusion due to the scalability requirement, the lack of a fusion center, and limited knowledge of the whole sensor network (more details in \cite{survey}). The authors of \cite{survey} surveyed both classical approaches and recent advances in multi-sensor data fusion for sensor networks. The authors of \cite{fusion1} reviewed the key theories and methodologies of distributed multi-sensor data fusion and discussed their advantages including graceful degradation, scalability, and interchangeability. \emph{Average consensus} was studied previously in distributed computing \cite{kalman-sensor} and for achieving consensus among agent values (a real number possibly representing its opinion or state). In \cite{cons}, a distributed consensus algorithm was developed for obtaining the averages of the node values over networks with large volume of data. The authors of \cite{dis_cons} proposed an asynchronous distributed average consensus algorithm to guarantee information-theoretic privacy in multi-agent systems. In \cite{consensus}, the authors provide a theoretical framework for analysis of consensus algorithms for multi-agent networked systems. In \cite{tracking1}, the authors developed a distributed consensus tracking filter to solve the target tracking problem. The authors in \cite{nedich} discussed algorithms for solving decentralized consensus optimization problems. None of the existing studies used average consensus algorithm for decentralized sensor data fusion for target tracking applications. Our study fills this knowledge gap.

In Chapter~\ref{chapter:formation}, we develop a novel decentralized UAV swarm formation control approach using Dec-MDP formation. In this problem, the goal is to optimize the UAV control decisions (e.g., bank angle and forward acceleration) in a decentralized manner such that the swarm forms a certain geometrical shape while avoiding collisions. We use dynamic programming principles to solve the decentralized swarm motion control problem. As most dynamic programming problems suffer from the curse of dimensionality, we adapt a fast heuristic approach called \emph{nominal belief-state optimization} (NBO) to solve the formation control problem approximately. We perform simulation studies to validate our control algorithms and compare their performance with centralized approaches for bench-marking the performance. 

In Chapter~\ref{chapter_target_tra}, we use decentralized Markov decision process (Dec-MDP) to solve UAV swarm control problem for multitarget tracking. We extend the above-mentioned ADP scheme NBO to this case study as well. We compare the performance (average target tracking error and average computation time) of our decentralized approaches to a centralized approach. 

In Chapter~\ref{chapter:consensus}, we develop a novel data fusion strategy for fusing information among a sensor network in a decentralized setting. These data fusion strategies can be easily extended to applications such as UAV networks or autonomous car networks. Specifically, we extend the \emph{average consensus algorithm} to perform decentralized data fusion while tracking a moving target via a sensor network. The sensor network is modeled by an undirected graph, which is assumed to be non-time varying. Each sensor generates a noisy measurement of the target state. The presence of an edge between the nodes or sensors means that the sensors are allowed to exchange information/messages for data fusion. In this study, we assume that each sensor maintains a local tracker (or tracking algorithm, e.g., Kalman filter), which updates its local target state estimate using the locally generated sensor measurements and the information it receives from its neighbors. We measure the performance of the above consensus algorithm using the performance metric \emph{average target tracking error} - the mean-squared error between the target state (ground truth) and the estimate. For bench-marking, we also implement the standard Bayesian data fusion approach and compare the performance of our approach with the Bayesian approach.


\section{Key Contributions}
\begin{itemize}
\item We formulate the UAV swarm formation control problem as a decentralized Markov decision process (Dec-MDP). 
\item We extend an approximate dynamic programming method called \emph{nominal belief-state optimization} (NBO) to solve the formation control problem. 
\item We perform simulation studies to validate the swarm formation control algorithms developed here. 
\item We perform numerical studies to quantify the impact of neighborhood threshold on average computation time and average pairwise distance.
\item One of the key contributions of this thesis is to induce cooperative behavior among the UAVs in the swarm as explained below: 
    \begin{itemize}
        \item Each UAV $i$ optimizes the control vector $[a_k^i, a_k^{nn}]$ at time $k$, where $a_k^i$ is the control vector for UAV $i$, and $a_k^{nn}$ is the control vector for its nearest neighbor. 
        \item Next, UAV $i$ discards the optimized controls for its neighbor and implements just its own controls $a_k^i$.
        \item Each UAV in the system implements the above approach.
    \end{itemize}

\item We extend the \emph{average consensus} algorithm \cite{cons} to track a moving target via a decentralized network of sensors. We compare the performance of this method against a standard benchmark method - \emph{decentralized Bayesian data fusion approach} \cite{fusion1}.    

\item We perform a numerical study to quantify the impact of various sensor network configurations (e.g., varying degrees of the nodes) on the performance of the \emph{average consensus} algorithm. 
\end{itemize}


\chapter{Decentralized Formation Shape Control of UAV Swarm}\label{chapter:formation}

\section{Introduction}

Unmanned aerial vehicle (UAV) swarm formation control has applications in various fields such as infrastructure inspection \cite{detection} and surveillance \cite{surveillance, surveillance1}, target tracking \cite{tracking}, and precision agriculture. The main objective in these application scenarios is to let the UAVs fly or hover in a certain geometrical formation, e.g., hover at locations lying on the surface of a sphere in a certain geographical region. There are methods existing in the literature to control UAV swarms using centralized methods \cite{ragi-tracking}, where there is a command center (centralized system) computing optimal motion commands for the UAVs. Centralized methods are comparatively easy to develop and implement, but when the swarm is large, the computational complexity for evaluating optimal motion commands is extremely high. This is because the computational complexity in optimizing the UAV motion commands grows exponentially with the number of UAVs in the system. We present a novel decentralized UAV swarm formation control approach. 

The word \emph{swarm} throughout this study refers to a collections of UAVs. Each UAV makes decisions on its local kinematic controls, i.e., bank angle and forward acceleration. All UAVs in the system are aware of the global objective, which is arriving at a position on a given geometrical surface in a geographical region. We call the geometrical region in which UAVs are supposed to arrive as \emph{formation shape}. Although these formation shapes could be three dimensional, we develop our control strategies in the context of 2-D \emph{formation shapes} for ease of implementation and to emphasize control strategies. The objective is to drive the swarm to the desired \textit{formation shape} in the shortest time possible while avoiding collisions among the UAVs.  

We pose this control problem as a \emph{decentralized Markov decision process} (Dec-MDP) and use a dynamic programming approach to solve the control problem. As dynamic programming problems suffer from the curse of dimensionality, we adapt a fast heuristic approach called \emph{nominal belief-state optimization} (NBO) to solve the formation control problem approximately.

The remaining parts of this Chapter are organized as follows. Section~\ref{sec:problem_spec} provides the problem specification. We formulate the problem using decentralized Markov decision process in Section~\ref{sec:problem_form} followed by the discussion on the NBO approach in Section~\ref{sec:NBO}. UAV motion model and kinematic equations are provided in Section~\ref{sec:motion_model}. In Section~\ref{sec:sim_result}, we discuss simulation results to evaluate the performance of our method. 

\section{Problem Specification}\label{sec:problem_spec}

\hspace{\parindent} \textbf{Unmanned aerial vehicles:} We assume UAV motion dynamics as described in \cite{ragi-path-planning}, where the motion controls are \emph{forward acceleration} and \emph{bank angle}. UAVs are allowed to hover at any location, i.e., the minimum speed limit on the UAVs is zero. 

\textbf{Communications and Sensing:} We assume that UAVs are equipped with sensing systems and wireless transceivers using which each UAV learns the exact location and the velocity of the nearest neighboring UAV. Our decentralized method requires only the state of the nearest neighbor to optimize the control commands of the local UAV.   

\textbf{Objective:} The goal is to control the swarm (optimizing control commands) in a decentralized manner such that the swarm arrives on a certain known geometrical surface in a certain region in the shortest time possible. The UAVs must complete this objective while avoiding collisions.  

\section{Problem Formulation}\label{sec:problem_form}
We formulate the swarm formation control problem as a decentralized Markov decision process (Dec-MDP). Dec-MDP is a mathematical formulation useful for modeling control problems for decentralized decision making. This formulation has the following advantages: 1) allows us to efficiently utilize the computing resources on-board all the UAVs, 2) requires less computational time compared to centralized approaches, 3) as UAVs are decentralized, point of failure of the entire mission is minimal, 4) decentralized approach provides robustness to addition or deletion of UAVs to the swarm, 5) UAVs do not need to rely on a central command center for evaluating optimal control commands. With Dec-MDP formulation, we can achieve the above features in our swarm control method. We define the key components of Dec-MDP as follows. Here, $k$ represents the discrete-time index. 

\subsection{Dec-MDP Ingredients}

\hspace{\parindent} \textbf{Agents/UAVs:} We assume there are $N$ UAVs in our system. The set of UAVs is given by $I = \{1,....,N\}$. Traditionally, this component is referred to as set of agents or set of independent decision makers. Here, an agent is a UAV. 

\textbf{States:} 
The state of the system $s_k$ includes the locations and velocities of all the UAVs in the system. 

\textbf{Actions:}
The actions are the controllable aspects of the system. We define action vector $a_k = (a_k^1,\ldots,a_k^N)$, where $a_k^i$ represents the action vector at UAV $i$, which includes the forward acceleration and the bank angle for the UAV. 

\textbf{State Transition Law:}
State transition law describes how the state evolves over time. Specifically, the transition law is a conditional probability distribution of the next state given the current state and the current control actions (assuming Markovian property holds). The transition law is given by $s_{k+1} \sim p_k(\cdot|s_k,a_k)$, where $p_k$ is the conditional probability distribution. Since the state of the system only includes the states of the UAVs, the state transition law is completely determined by the kinematic equations of the UAVs (discussed in the next section). In other words, the transition law is given by $s_{k+1}^i = \psi(s_k^i,a_k^i) + w_k^i, i=1,\ldots,N$, where $s_k^i$ represents the state of the $i$th UAV and $a_k^i$ indicates the local kinematic controls (forward acceleration and bank angle) of $i$th UAV, $\psi$ represents the kinematic motion model as discussed in Section~\ref{sec:motion_model}, and $w_k^i$ represents noise, which is modeled as a zero-mean Gaussian random variable.

\textbf{Cost Function:}
The cost function $C(s_k,a_k)$ deals with cost of being in a given state $s_k$ and performing actions $a_k$. Here, $s_k$ represents the global state, i.e., the state of all the UAVs in the system. Since the problem is decentralized, each UAV only has access to its local state and the state of the nearest neighboring UAV. Let $b_k^i = (s_k^i,s_k^{nn})$ represent that local system state at UAV $i$, where $s_k^{nn}$ is the state of the nearest neighboring UAV, and $nn \in I \backslash \{i\}$. 

Let $d^i$ is the destination location UAV $i$ must reach, and $d_{\mathrm{coll,thresh}}$ is the distance between the UAVs below which the UAVs are considered to be at the risk of collision. We now define the local cost function for UAV $i$ as follows:

\begin{equation} \begin{split}
 c(b_k^i,a_k^i,a_k^{nn}) &= w_1 \left[\mathrm{dist}(s_k^{i,\mathrm{pos}},d^i) + \mathrm{dist(s_k^{nn,\mathrm{pos}},d^{nn})}\right] \\ & + w_2 \left[\mathrm{dist}(s_k^i,s_k^{nn})^{-1} \mathbb{I}(\mathrm{dist}\left(s_k^i,s_k^{nn}) < d_{\mathrm{coll,thresh}} \right)  \right] \end{split} \end{equation}

where $s_k^{i,\mathrm{pos}}$ represents the location of the $i$th UAV, $w_1$ and $w_2$ are weighting parameters, $\mathrm{dist}(a,b)$ represents the distance between locations $a$ and $b$, and $\mathbb{I}(a)$ is the indicator function, i.e., $\mathbb{I}(a) = 1$ if the argument $a$ is true and $0$ otherwise.

By minimizing the above cost function, each UAV optimizes its own control commands and that of its neighbor, but only implement its own local control commands and discards the commands optimizes for its neighbor. The first part of the cost function lets the UAV reach its destination, while the second part minimizes the risk of collisions between UAVs. 

The Dec-MDP starts at an initial random state $s_0$ and the state of the system evolves according to the state-transition law and the control commands applied at each UAV. The overall objective is to optimize the control commands at each UAV $i$ such that the expected cumulative local cost over a horizon $H$ (shown below) is minimized. 
\begin{equation}\label{eq:optim}
\min_{\{a_k^i,a_k^{nn}\}, k=0,\ldots,H-1} \text{E} \left[\sum_{k=0}^{H-1} c(b_k^i,a_k^i,a_k^{nn}) \middle\vert b_0^i\right]
\end{equation}
where $b_0^i$ is the initial local state at UAV $i$, and the expectation $E[\cdot]$ is over the stochastic evolution of the local state over time (due to the random variables present in the UAV kinematic equations). 

\section{NBO Approach to Solve Dec-MDP}\label{sec:NBO}
It is well know in the literature that solving Equation~\ref{eq:optim} exactly is computationally prohibitive and not practical. For this reason, we extend a heuristic approach called \emph{nominal belief-state optimization} (NBO) \cite{ragi-path-planning}. As discussed in the previous section, we let a UAV optimize its own and its nearest neighbor's kinematic controls over the time horizon $H$. Once the UAV calculates local controls for itself and its neighbors, the UAV implement its own controls and discards its neighbors controls at each time step. Since obtaining the expectation in Equation~\ref{eq:optim} exactly is not tractable, the NBO approach approximates this expectation by assuming that all the future random variables (over which the expectation is supposed to be evaluated) assume the nominal values, i.e., the mean values. Since we model the above-mentioned random variable as zero-mean Gaussian, the nominal values are simply zeros. In summary, the NBO approach approximates the cumulative cost function in Equation~\ref{eq:optim} by replacing the expectation with the random trajectory of the states over time by a sequence of states obtained by replacing future random variables with zeros. 

In the NBO method, the objective function at agent $i$ is approximated as follows:
\begin{equation*}
J(b_0^i)\approx \sum_{k=0}^{H-1} c(\hat b_k^i,a_k^i,a_k^{nn}),
\end{equation*}
where $\hat b_1^i,\hat b_2^i,\ldots, \hat b_{H-1}^i$ is a \emph{nominal} local state sequence.

\section{UAV Motion Model}\label{sec:motion_model}
The state of the $i$th UAV at time $k$ is given by $s_k^i = \left(p_k^i,q_k^i,V_k^i,\theta_k^i\right),$ where $(p_k^i,q_k^i)$ represents the position coordinates, $V_k^i$ represents the speed, and $\theta_k^i$ represents the heading angle. The kinematic control action for UAV $i$ is given by $a_k^i = (f_k^i,\phi_k^i)$, where $f_k^i$ is the forward acceleration and $\phi_k^i$ is the bank angle of the UAV. The kinematic equations of the UAV motion \cite{ragi-path-planning} are as follows: \begin{equation*}
\begin{aligned}
V_{k+1}^i &= \left[V_k^i + f_k^iT\right]_{V_{\min}}^{V_{\max}} + w_k^{i,\mathrm{speed}}\\
\theta_{k+1}^i &= \theta_k^i + (gT\tan(\phi_k^i)/V_{k}^i) + w_k^{i,\mathrm{heading}},\\
p_{k+1}^i &= p_k^i + V_{k}^iT\cos(\theta_{k}^i) + w_k^{i,\mathrm{xpos}},\\
q_{k+1}^i &=q_k^i + V_{k}^iT\sin(\theta_{k}^i) + w_k^{i,\mathrm{ypos}},
\end{aligned}
\end{equation*}
where $[v]_{V_{\min}}^{V_{\max}}=\max\left\{V_{\min},\min(V_{\max},v)\right\}$, $V_{\min}$ and $V_{\max}$ are the minimum and the maximum limits of each UAV, $g$ is the acceleration due to gravity, $T$ is the length of the time step, and $w_k^{i,\mathrm{speed}}, w_k^{i,\mathrm{heading}}, w_k^{i,\mathrm{xpos}}, w_k^{i,\mathrm{ypos}}$ are the zero-mean Gaussian random variables. 

\section{Simulation Results}\label{sec:sim_result}
We assume that each UAV has its own on-board computer to compute the local optimal control decisions. We implement the above-discussed NBO approach to solve the swarm control problem in MATLAB. We test our methods with three formation shapes - a circular shape, a rectangular shape, and a square shape. The UAVs are aware of the shape dimensions and the exact location of shape. Each UAV randomly picks a location on the formation shape, and uses the NBO approach to arrive at this location. We use MATLAB's \emph{fmincon} to solve the NBO optimization problem. Here, we set the horizon length to $H=7$ time steps. 

We define the following metrics to measure the performance of our formation control approach: 1) $T_c$ - \emph{average computation time to evaluate the optimal control commands} and 2) $T_f$: \emph{time taken for the swarm to arrive on the formation shape}. As a benchmark method, we use a centralized approach to solve the above-discussed swarm formation control problem. In other words, we use a single NBO algorithm, which optimizes the motion control  commands for all the UAVs together based on the global state of the system. We implement this centralized algorithm in MATLAB.


\begin{figure}%
    \centering
    \subfloat[Circular formation]
    {\includegraphics[width= 0.35\columnwidth, trim = 130 180 80 150,clip]{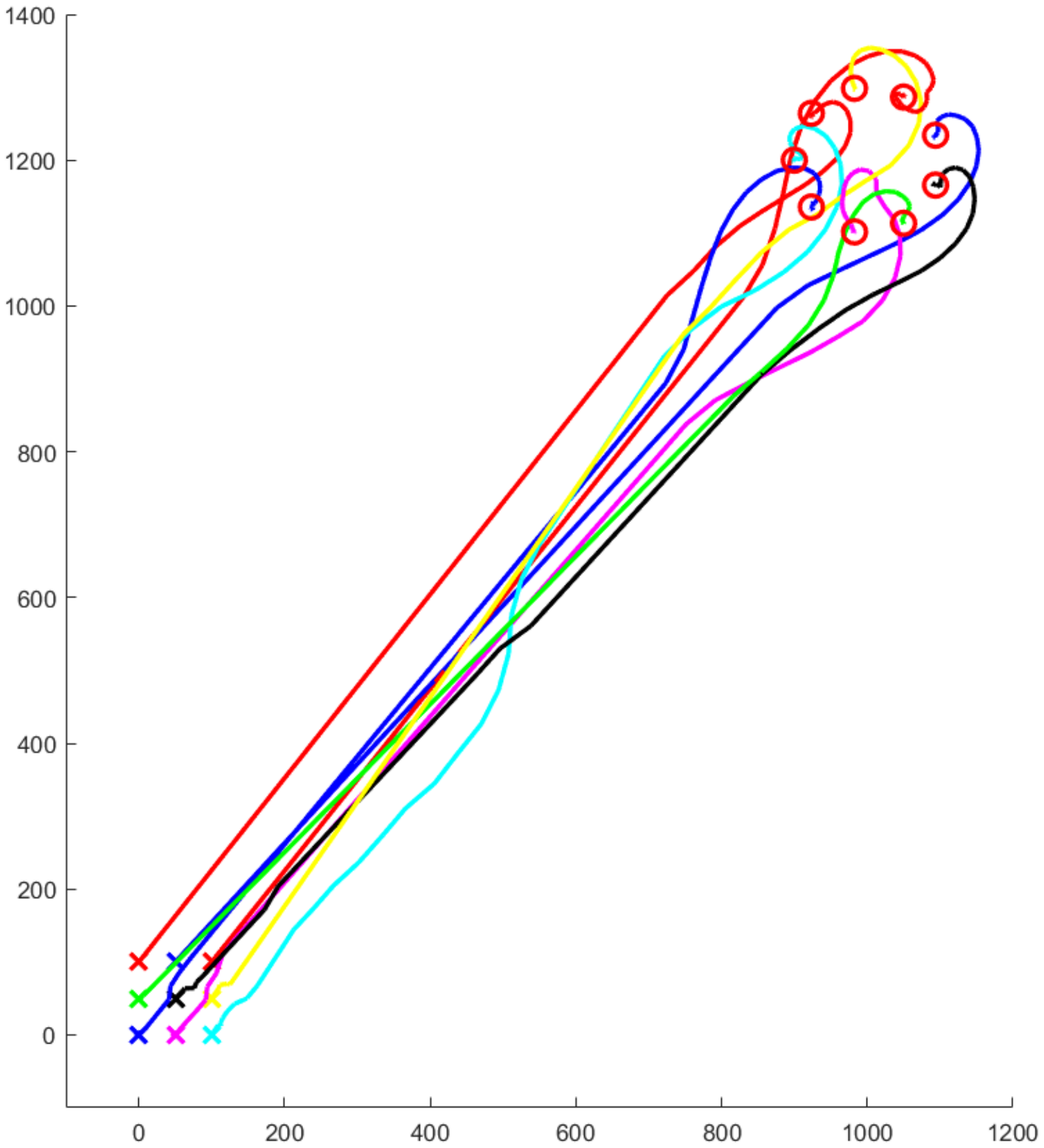}\label{fig: circular}}
    \quad
    \subfloat[Rectangle formation]
    {\includegraphics[width= 0.25\columnwidth, trim = 190 180 180 150,clip]{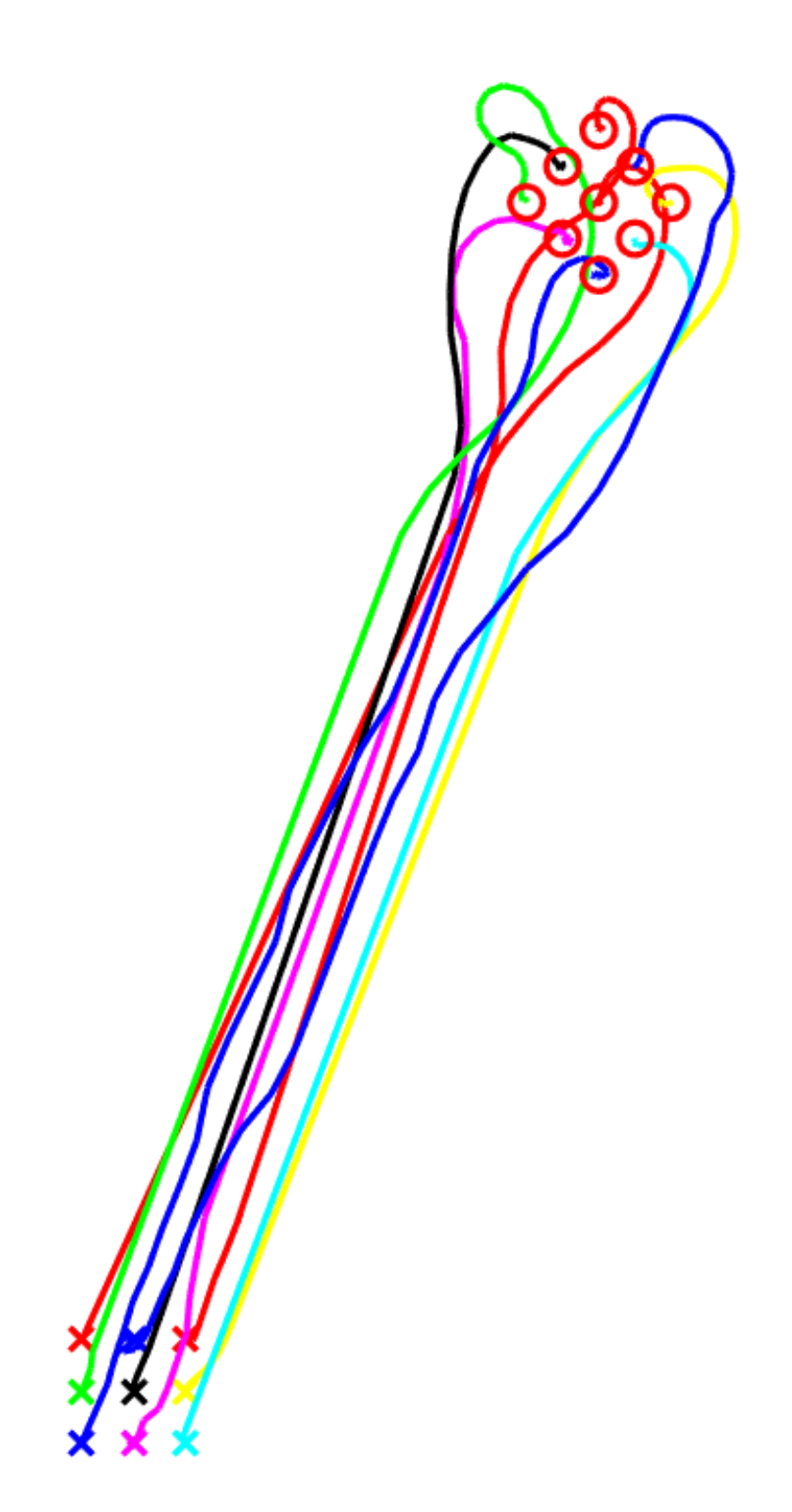}\label{fig: rect}}
    \qquad
    \subfloat[Square formation]
    {\includegraphics[width= 0.25\columnwidth, trim = 190 180 180 150,clip]{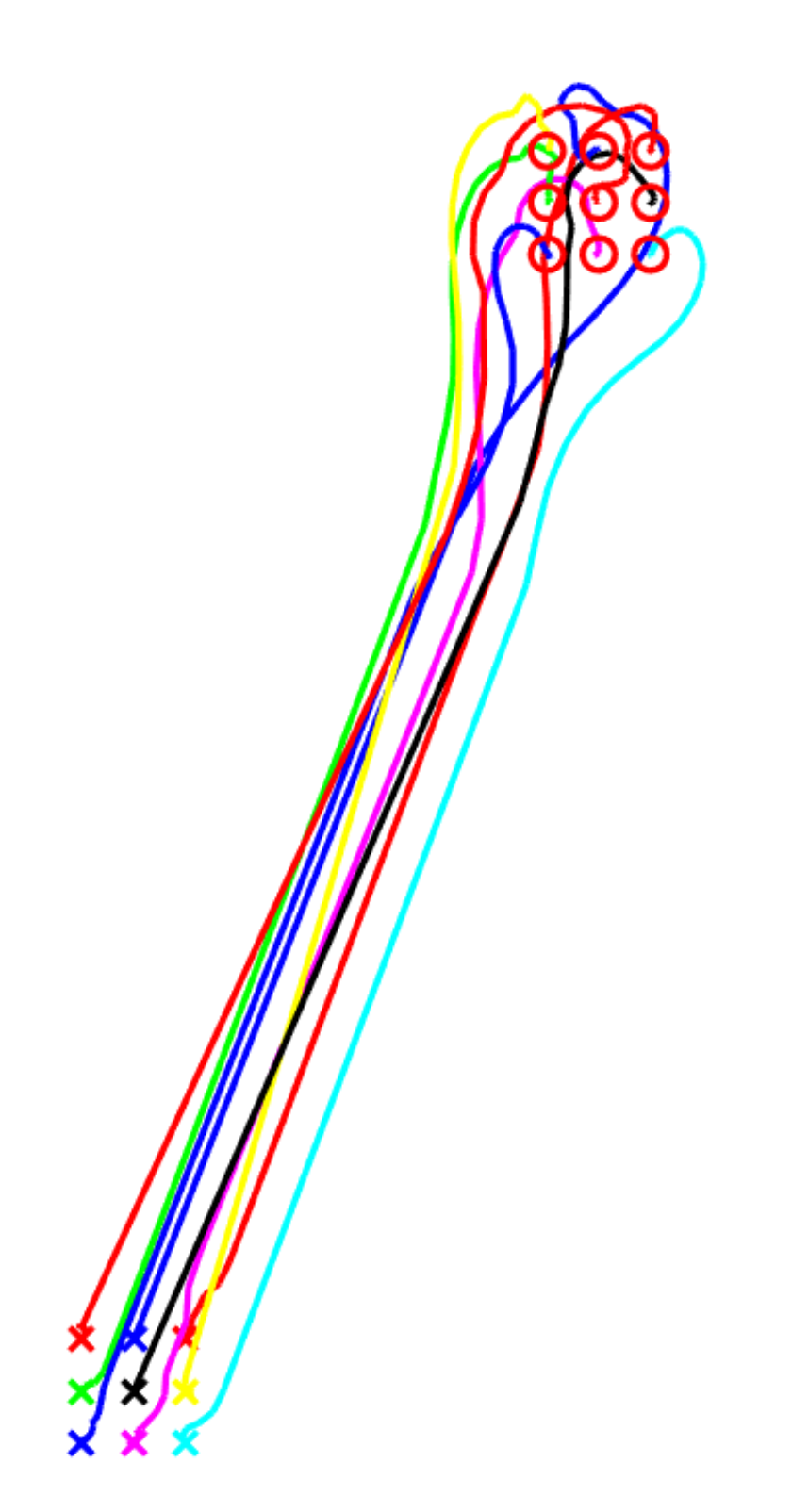}\label{fig: sq}}
    \caption{9 UAVs converging to the formation shapes using the \emph{Dec-MDP approach}}
    \label{fig:example}%
\end{figure}



\begin{figure}
\centering{\includegraphics[width= 0.8\columnwidth, trim = 10 150 20 150,clip]{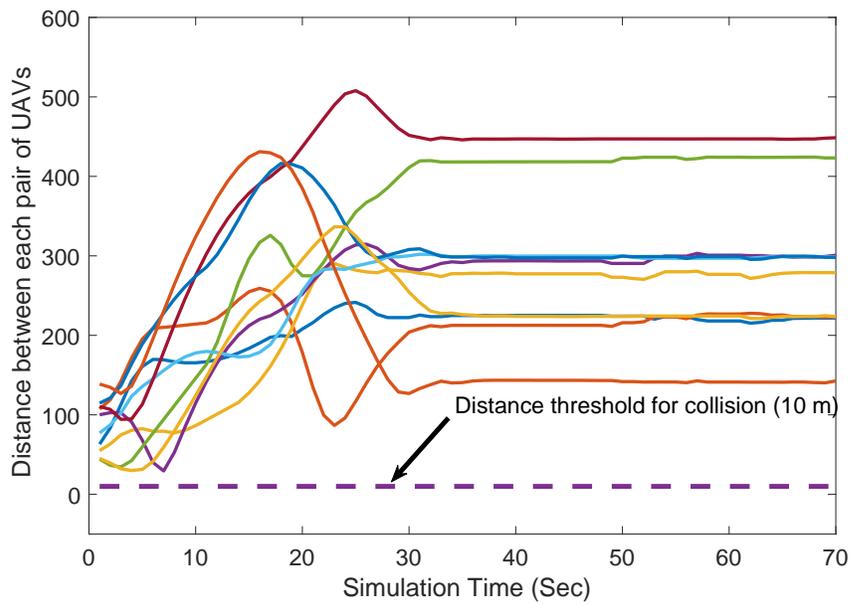}}
\caption{Distance between each pair of UAVs}
\label{fig:distance_UAVs_dec_2}
\end{figure}

We implement the Dec-MDP approach with a circular formation shape, a rectangular formation shape, and a square formation shape. The resulting swarm motion is shown in Figures~\ref{fig: circular}, \ref{fig: rect}, and \ref{fig: sq} respectively. For the scenario in Figure~\ref{fig: circular}, we also plot the distance between every pair of UAVs in the swarm as shown in Figure~\ref{fig:distance_UAVs_dec_2}. Here, we assume that there is a collision risk between a pair of UAVs when the distance between them is less than 10 m.  Clearly, the Figures~\ref{fig:example}, and \ref{fig:distance_UAVs_dec_2} demonstrate that our decentralized algorithm drives the swarm to the destination while successfully avoiding collisions between the UAVs.

\begin{figure}
\centering{\includegraphics[width= 0.8\columnwidth, trim = 10 150 20 150,clip]{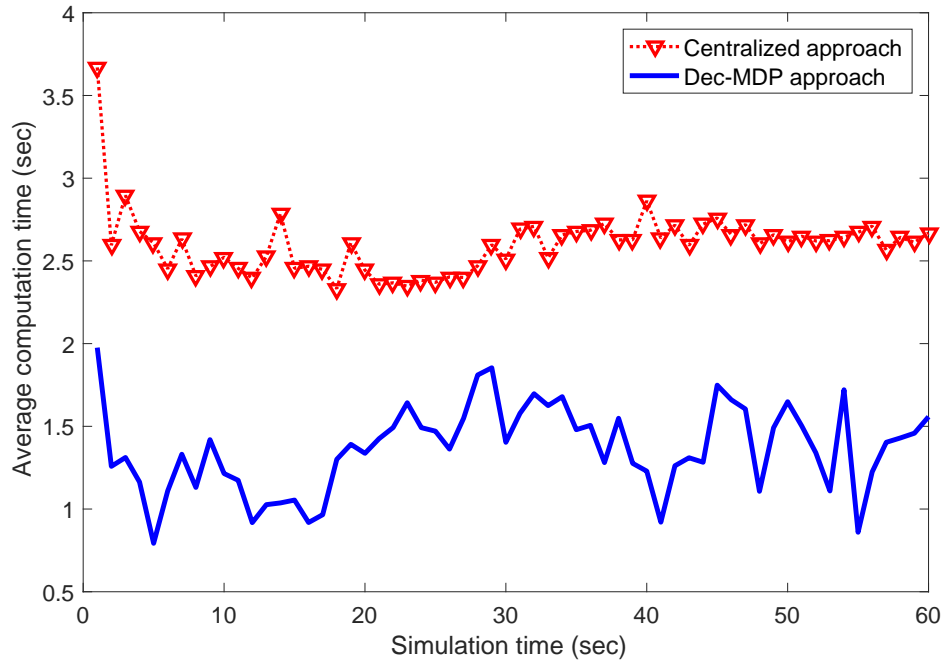}}
\caption{Computation time ($T_c$): centralized vs decentralized method}
\label{fig:comp_time}
\end{figure}

We calculate the $T_c$ and $T_f$ values for both the centralized and the decentralized algorithms for 9 UAVs. Figure~\ref{fig:comp_time} and Table I clearly demonstrates that our decentralized method significantly outperforms the centralized method with respect to both the metrics $T_c$ and $T_f$.

\begin{table}
\begin{center}
\begin{tabular}{|c|c|c|}
\hline
  & Dec-MDP & Centralized  \\
\hline
$T_f (sec)$ & 16.7 & 25.98 \\
\hline
\end{tabular}
\label{tab:table}
\\[10pt]
\caption{Average time taken by the swarm to arrive at the formation shape.}
\label{table}
\end{center}
\end{table}

\begin{figure}
\centering{\includegraphics[width= 0.78\columnwidth, trim = 120 235 120 250,clip]{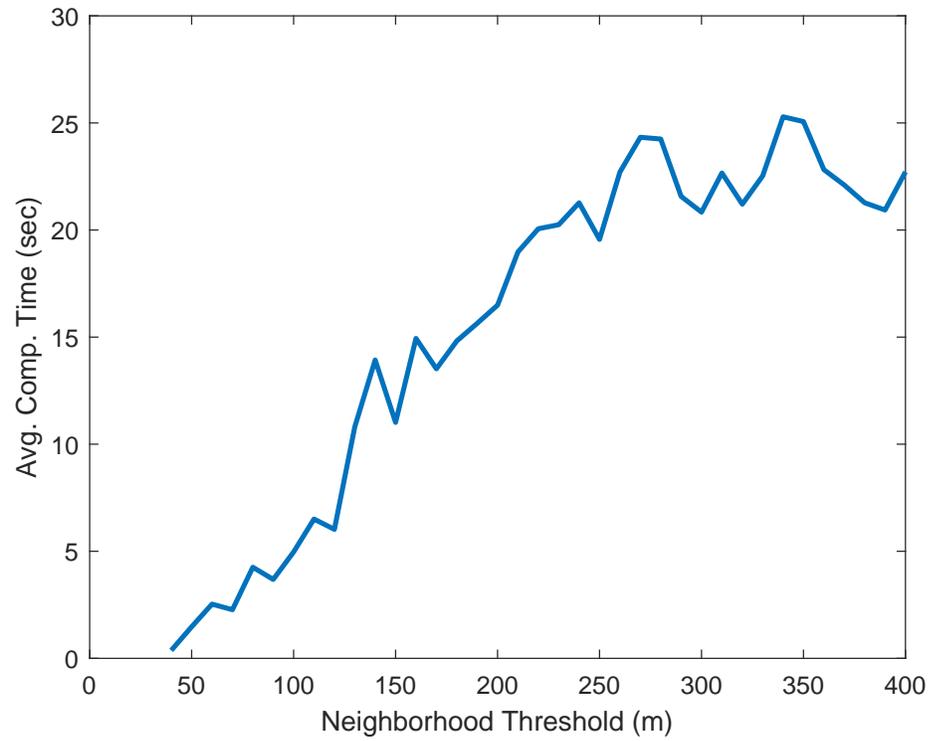}}
\caption{Average computation time with respect to neighborhood threshold}
\label{fig: computation_time}
\end{figure}

\begin{figure}[!htbp]
\centering{\includegraphics[width= 0.78\columnwidth, trim = 120 235 120 250,clip]{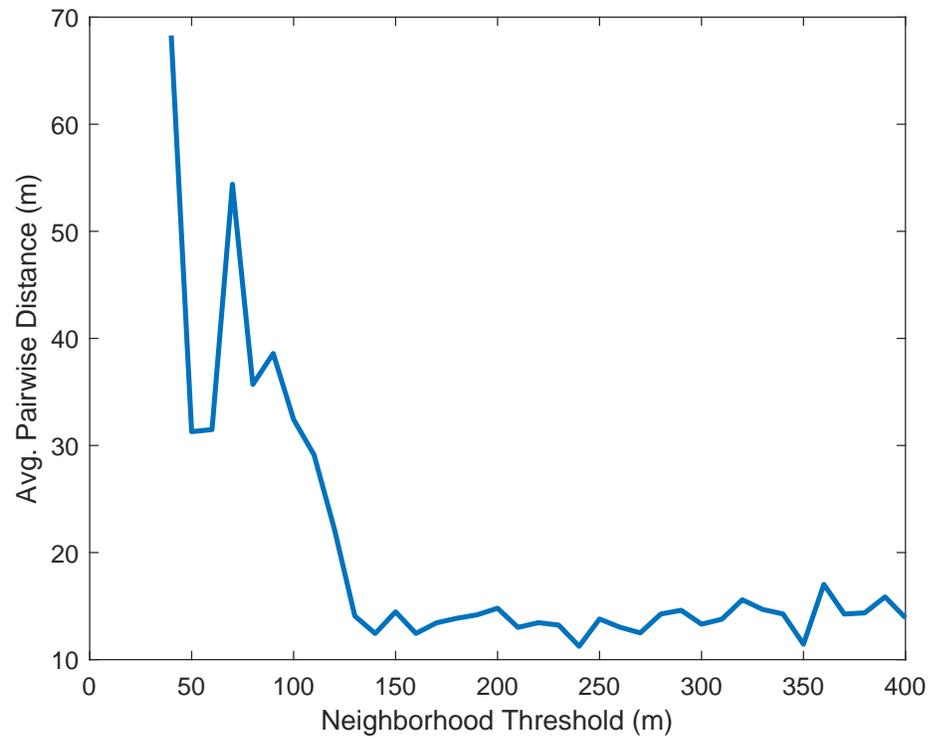}}
\caption{Average pairwise distance with respect to neighborhood threshold}
\label{fig: pairwise_dist}
\end{figure}

We now compute average computation time and average pairwise distance with respect to neighborhood threshold where each UAV communicates with other UAVs within the radius of neighborhood threshold. If neighborhood threshold is infinity, a UAV can communicate with all other UAVs in the swarm. UAVs optimize its decision together with neighbors which depends on neighborhood threshold and implement its own control. We expect that with the increase of neighborhood threshold, average computation time rises and after certain neighborhood threshold, average computation time saturates. Figure~\ref{fig: computation_time} shows average computation time rise until neighborhood threshold reach 240 m and then waves between 20 to 25 sec.

We also expect that with the increase of neighborhood threshold, average pairwise distance drops. The reason we are interested in analyzing average pairwise distance is, we expect the swarm to be as closely as possible while avoiding collision between UAVs. Small average pairwise distance allows the swarm to be more cooperative while saving battery life as communication distance depends on distance between UAVs. Figures~\ref{fig: pairwise_dist} and \ref{fig: computation_time} suggest that neighborhood threshold more than 130 m allows UAVs to stay closely in the swarm with reasonable computation cost.




\chapter{Target Tracking with Decentralized UAV Swarm}\label{chapter_target_tra}
\section{Introduction}
In this chapter, we extend the Dec-MDP framework for a UAV swarm control problem for single and multitarget tracking applications. Target tracking using UAV swarms is a well studied problem in the literature owing to their applications in surveillance and monitoring. For instance, \cite{zhan2005, hao2016} studied centralized UAV control methods for target tracking. The authors of \cite{ragi-path-planning} developed a UAV control problem as a partially observable Markov decision process (POMDP) for a target tracking application. As mentioned in the earlier chapters, centralized control methods are computationally expensive especially when the swarm is large. Specifically, the computational complexity grows exponentially with the number of UAVs in the swarm. To tackle this challenge, several research studies were carried out previously. For example, the authors of \cite{wei2014} presented a distributed multi-UAV target search algorithm for search, tasking, and tracking ground targets. They combine urban road map and target detection probability map information for UAV guidance and control. However, they did not formulate a decision theoretic approach for UAV guidance. 

The authors of \cite{kim2016} used a graph-theoretic approach to guide the UAVs while tracking a target, and used potential field-based approaches to avoid collisions. In this paper, the authors designed a decentralized controller for agents which is a modification of navigation function developed in \cite{Kan2012}. The authors of \cite{ragi-dec-pomdp} posed a UAV control problem as a \textit{decentralized partially observable Markov decision process} or Dec-POMDP.  

Inspired from these efforts, we develop a decentralized UAV swarm control strategies using a decision theoretic framework called decentralized Markov decision process (Dec-MDP). We develop these methods in the context of single and multitarget tracking applications. Typically, these Dec-MDP problems are studied using dynamic programming (DP) \cite{dp_Bertsekas}. DP problems are computationally hard and not tractable. So, a plethora of approximation methods exist in the literature called approximate dynamic programs (ADPs). A survey of these ADP schemes can be found in \cite{chong2009}. In our study, we extend an ADP method called \emph{nominal belief-state optimization} (NBO) \cite{miller2009,ragi-path-planning}, which is computationally the most efficient compared to other ADP schemes in the literature.   

\section{Problem Specification}\label{spec_track}
We assume the targets move in a 2D plane for simplicity. These methods can be easily extended to 3D. We use a 2D motion model for the UAVs assuming the altitude of the UAVs to be constant; the kinematic equations that drive the UAVs are discussed in Section~\ref{sec:motion_model}. The motion of the UAVs is controlled by the forward acceleration and the bank angle. We assume that the UAVs are equipped with sensors on-board that generate the position coordinates of the targets, albeit these measurements are corrupted by random noise. Our aim is to develop a decentralized control algorithm that runs on each UAV and performs the following tasks: collects the target measurements, constructs the target state estimated, and computes the control commands for maximizing the target tracking performance. The target measurement error at a UAV depends on the position of the UAV and the target. The objective is to minimize the target tracking error measured as the mean-squared error between the target state and its estimate. We assume that the total number of targets are less than or equal to the total number of UAVs in this study.


\section{Dec-MDP Formulation and NBO approximation}\label{form_track}
The benefits of using a Dec-MDP formulation is explained in Section~\ref{sec:problem_form}, which hold here as well. Before we define the key elements of Dec-MDP for the current case study, we define the state of the system as follow. 

Suppose $k$ represents the discrete time index. The system state at time $k$ is given by $x_k$ = $(s_k, \chi_k, \xi_k, \mathbf{P}_k)$ where $s_k$ represents the state of the UAVs which includes location and velocity of all the UAVs in the system, $\chi_k$ represents the target state including the locations, velocities, and accelerations of the targets. $(\xi_k, \mathbf{P}_k)$ represents the tracker state, which is the state of the tracking algorithm (in our study we use the standard Kalman filter \cite{Blackman,Bar2001}), where $\xi_k$ is the posterior mean vector and $\mathbf{P}_k$ is the posterior co-variance matrix. 

The target state is not fully observable; we infer the target state via Kalman filter using the noisy measurements and the target motion model. Since one of the state variables is not observable, if we were to use this state definition as the state of the system, then the system dynamics cannot be modeled via Dec-MDP since the state in Dec-MDP is assumed to be observable. To formulate this problem as a Dec-MDP, we instead use the ``belief state'' as the state of the system. Belief state is the posterior distribution over the state space. Let $B_k$ represents the belief state at time $k$ given by $(B_k^s,B_k^{\chi},B_k^{\xi},B_k^{P})$, where
\[
\begin{aligned}
B_k^s &= \delta(s-s_k)\\
B_k^{\xi} &= \delta(\xi-\xi_k)\\
B_k^{P} &= \delta(P-P_k)\\
\end{aligned}
\]
and $B_k^{\chi} = \mathcal{N}(\xi_k,P_k)$. Since the UAV and tarcker states are fully observable, the corresponding belief states are represented by the delta functions as shown above. The target belief state is given by the Gaussian distribution with mean and covariance matrix given by the elements in the tracker state.    

\subsection{Dec-MDP Elements}

\textbf{States:} 
The state at time $k$ is given by $B_k$ = $(B_k^s, B_k^{\chi}, B_k^{\xi}, B_k^{P})$ as discussed above. 

\textbf{Actions:} 
The action vector $a_k = (a_k^1,\ldots,a_k^N)$, where $a_k^i$ represents the control decisions at UAV $i$, which includes the forward acceleration and the bank angle for the UAV. 

\textbf{State Transition Law:}
Given the current state and the control action, the transition law is the conditional probability distribution of the next state. The transition law is given by $B_{k+1} \sim p(\cdot|B_k,a_k)$, where $p$ is the conditional probability distribution. The state transition law is determined by the motion model described in the Section~\ref{sec:motion_model}. 
We model the target's motion using constant velocity model as shown below. 
$$
\chi_{k+1} = \mathbf{F}_k\chi_{k} + v_k, \: v_k \sim \mc{N}(0,\mathbf{V}_k), 
$$
where $\mathbf{F}_k$ is the target motion model (same for all targets), and $\mathbf{V}_k$ is the covariance matrix of the additive process noise \cite{Bar2001,Blackman}. Finally, the tracker state evolves according to the Kalman filter equations \cite{Blackman, Bar1998}. In essence, the Kalman filter equations capture the state-transition law for the target state and the tracker state.  

\textbf{Cost Function:}
Suppose $B_k^i$ is the local state of the system at agent $i$. The cost function defines the cost of taking an action in a given state at agent $i$. We use the mean-squared error between the tracks and the targets as the cost function at agent $i$ given by:
\begin{equation}\label{cost_func}
c(B_k^i, a_k^i, a_k^{nn}) = w_1 \mathrm{Tr}(P_k^i) + w_2\left[\mathrm{dist}(s_k^{i,\mathrm{pos}},s_k^{nn,\mathrm{pos}})^{-1} \mathbb{I}(\mathrm{dist}\left(s_k^{i,\mathrm{pos}},s_k^{nn,\mathrm{pos}}) < d_{\mathrm{coll,thresh}} \right)  \right] \end{equation}

where $P_k^i$ is the target state's posterior covariance matrix at agent $i$, $a_k^{nn}$ is the set of actions of the nearest neighbor (same as the definition in the previous chapter), $s_k^{nn}$ represents the state of the nearest neighboring UAV, $s_k^{i,\mathrm{pos}}$ represents the location of the $i$th UAV, $w_1$ and $w_2$ are weighting parameters, $\mathrm{dist}(a,b)$ represents the distance between locations $a$ and $b$, and $\mathbb{I}(a)$ is the indicator function, i.e., $\mathbb{I}(a) = 1$ if the argument $a$ is true and $0$ otherwise. The first component in the cost function captures the target tracking performance and the second component captures the penalty for collision. 

\subsection{Optimal Policy}
To minimize the cost function in Equation~\ref{cost_func}, each UAV optimizes its own actions and its nearest neighbors' actions. After this step, the UAV implements its own actions and discards the neighbors' actions. 

The objective is to choose control commands for a UAV $i$ over a time horizon $H$ such that expected cumulative cost is minimized. The cumulative cost at $i$ over time horizon $H$ can be written as follows. 
\begin{equation}\label{eq:optim_2}
J_H =  \text{E} \left[\sum_{k=0}^{H-1} c(B_k^i,a_k^i,a_k^{nn})\right]
\end{equation}

Equation~\ref{eq:optim_2} is hard to solve exactly due to its exponential computational complexity in worst-case. We extend a heuristic approach called \emph{nominal belief-state optimization} (NBO) to solve the Equation~\ref{eq:optim_2} approximately, which is described in Section~\ref{sec:NBO}. The objective function at agent $i$ is then approximated as follows:
\begin{equation}\label{eq:optim_reqrite1}
J_H \approx \sum_{k=0}^{H-1} c(\hat{B}_k^i,a_k^i,a_k^{nn})
\end{equation}
where $\hat{B}_0^i$, $\hat{B}_1^i$, \ldots, $\hat{B}_{H-1}^i$ is the nominal local belief-state sequence obtained from the NBO approach.  

\section{Simulation Results}\label{result_track}
In this part, we evaluate performance of our methods for single target and multitarget tracking with 5 UAVs. We compare the performance of our Dec-MDP method with a centralized approach as a benchmark. We find out average target tracking error for both single target and multitarget tracking in Dec-MDP setting and compare with centralized method. Average target tracking error is evaluated each time step over all targets and all sensors. We also find out average computation time taken by our decentralized approach and compare with the centralized approach. 

\begin{figure}[!htbp]
\centering{\includegraphics[width= \columnwidth, trim = 200 50 160 50,clip]{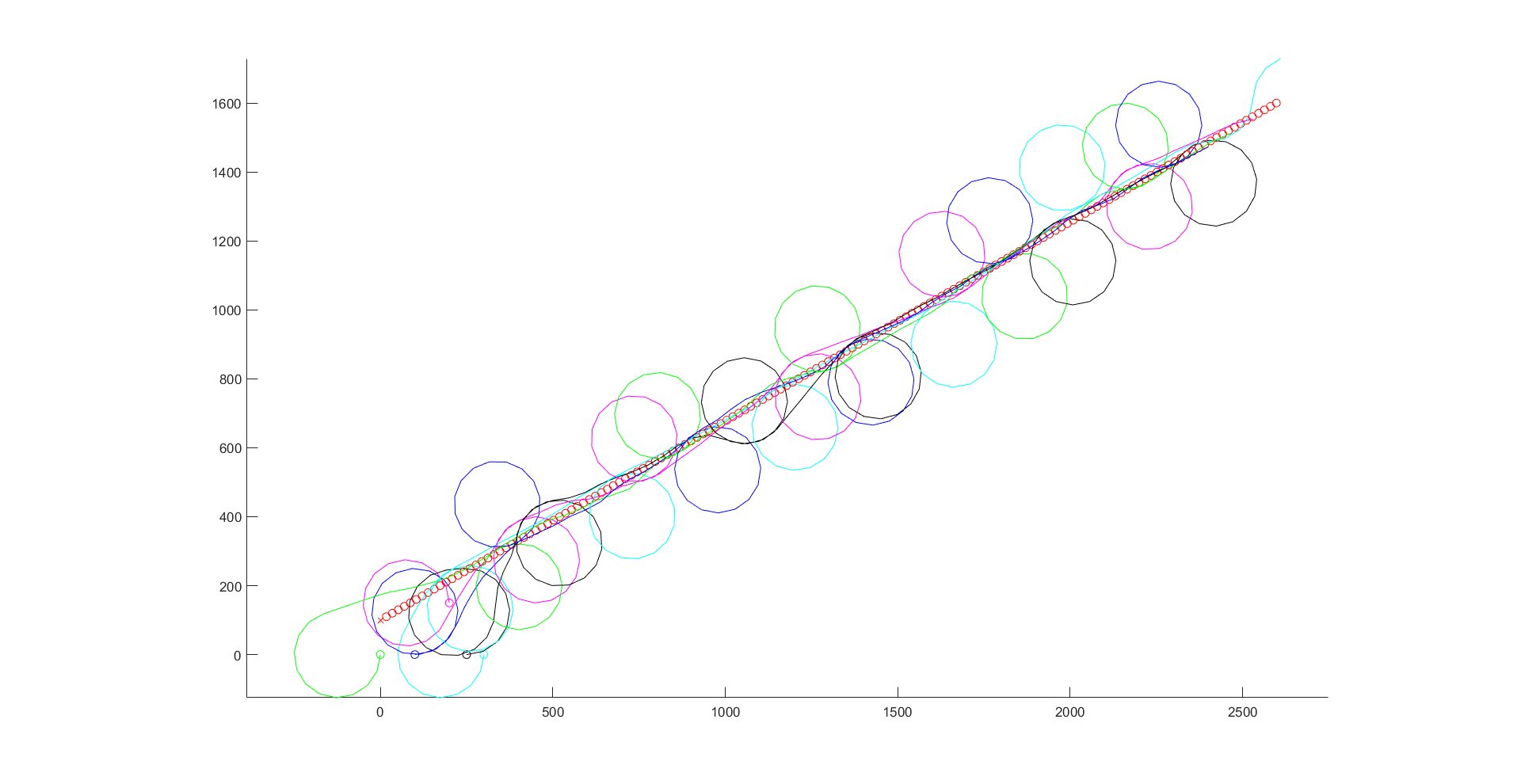}}
\caption{5 UAVs tracking a single target}
\label{fig: single_target}
\end{figure}

\begin{figure}[!htbp]
\centering{\includegraphics[width= \columnwidth, trim = 200 50 160 50,clip]{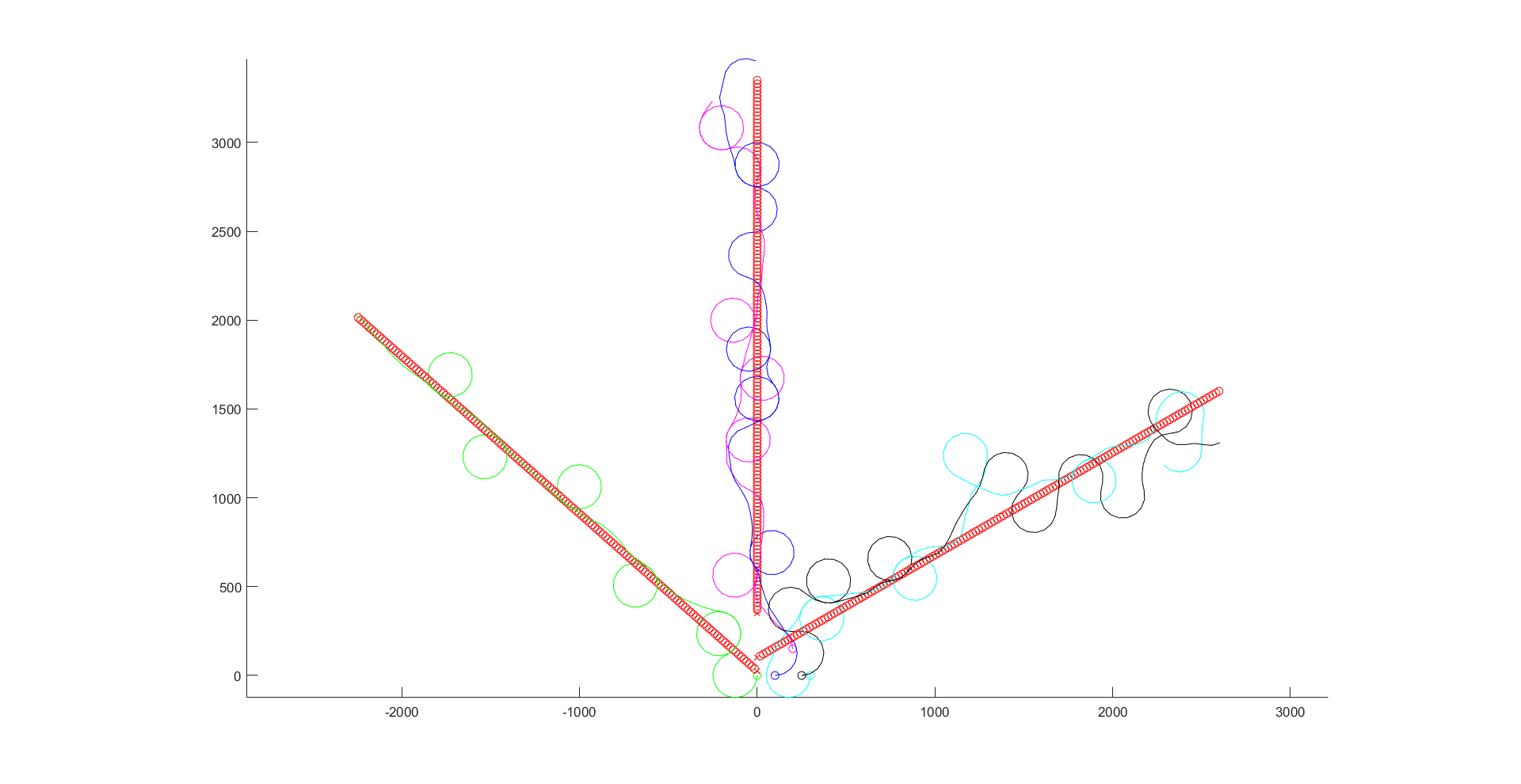}}
\caption{5 UAVs tracking 3 targets}
\label{fig: multitarget}
\end{figure}

Figures~\ref{fig: single_target} and \ref{fig: multitarget} illustrate the simulation of 5 UAVs tracking a single and 3 targets respectively. Targets start from near origin for both figures and moves with a constant speed. The UAVs also start near the origin and move according to kinematic control obtained from the \textit{fmincon}. For the single target scenario, all UAVs are tracking the target in the Figure~\ref{fig: single_target}. For the multitarget tracking, the UAVs are assigned to a target. UAVs track the associated target showed in Figure~\ref{fig: multitarget}. We run the simulation for 150 discrete time steps to measure the performance (average target tracking error and average computation time over all targets and all sensors) of our approach.

\begin{figure}[!htbp]
\centering{\includegraphics[width= 0.7\columnwidth, trim = 110 235 120 240,clip]{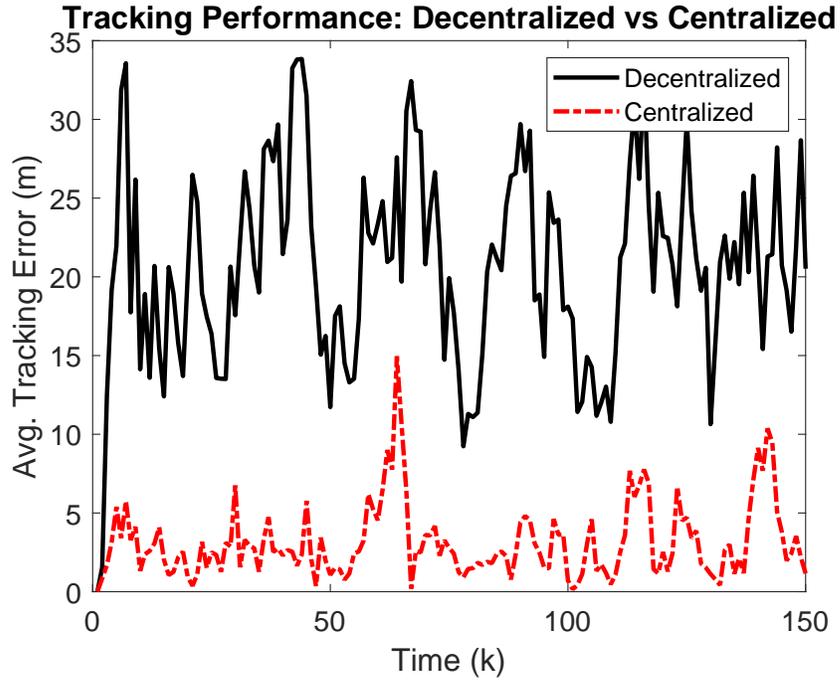}}
\caption{Average target tracking error: single target}
\label{fig: error}
\end{figure}

\begin{figure}[!htbp]
\centering{\includegraphics[width= 0.7\columnwidth, trim = 105 235 120 240,clip]{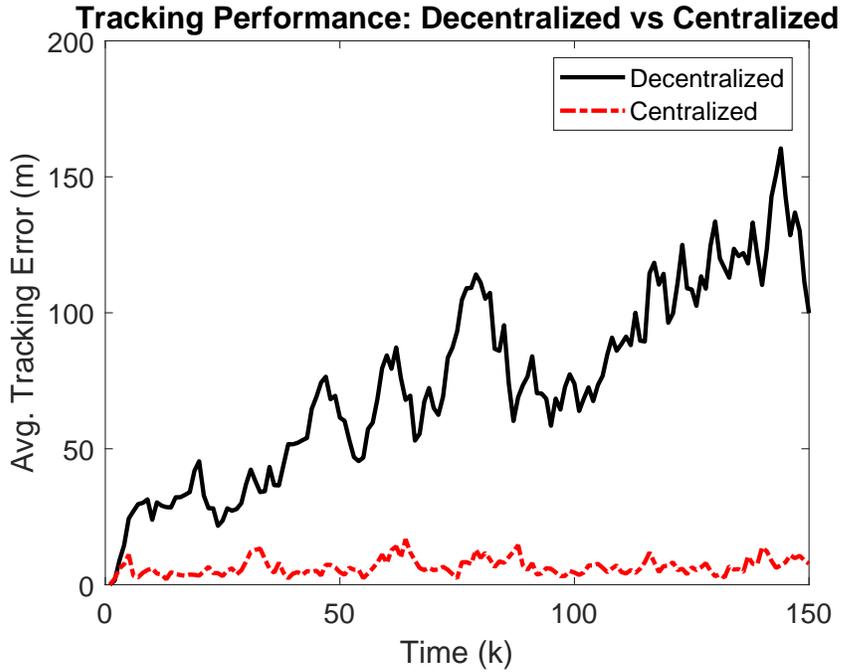}}
\caption{Average target tracking error: multitarget}
\label{fig: error_multi}
\end{figure}

\begin{figure}[!htbp]
\centering{\includegraphics[width= 0.7\columnwidth, trim = 110 235 120 240,clip]{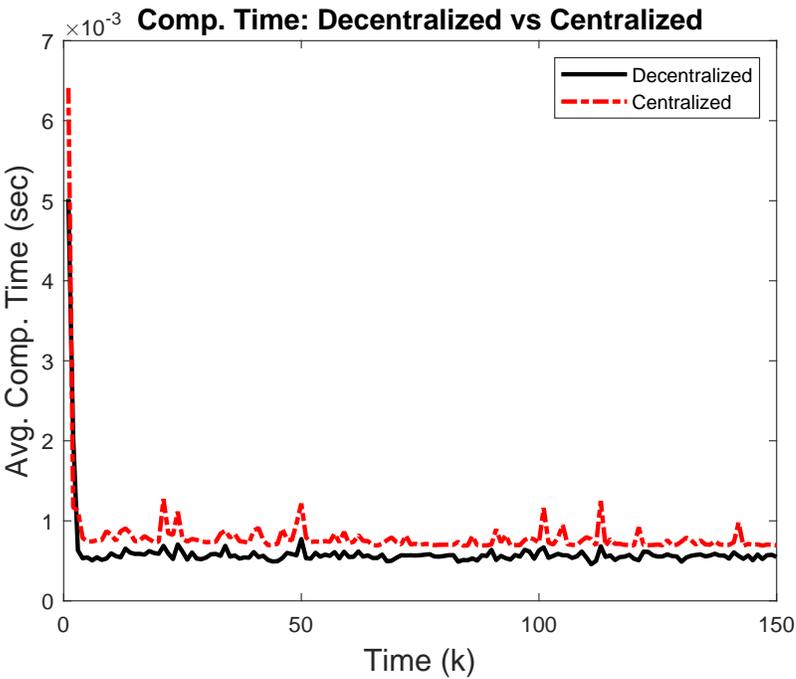}}
\caption{Average computation time: single target}
\label{fig: time}
\end{figure}

\begin{figure}[!htbp]
\centering{\includegraphics[width= 0.7\columnwidth, trim = 105 235 120 240,clip]{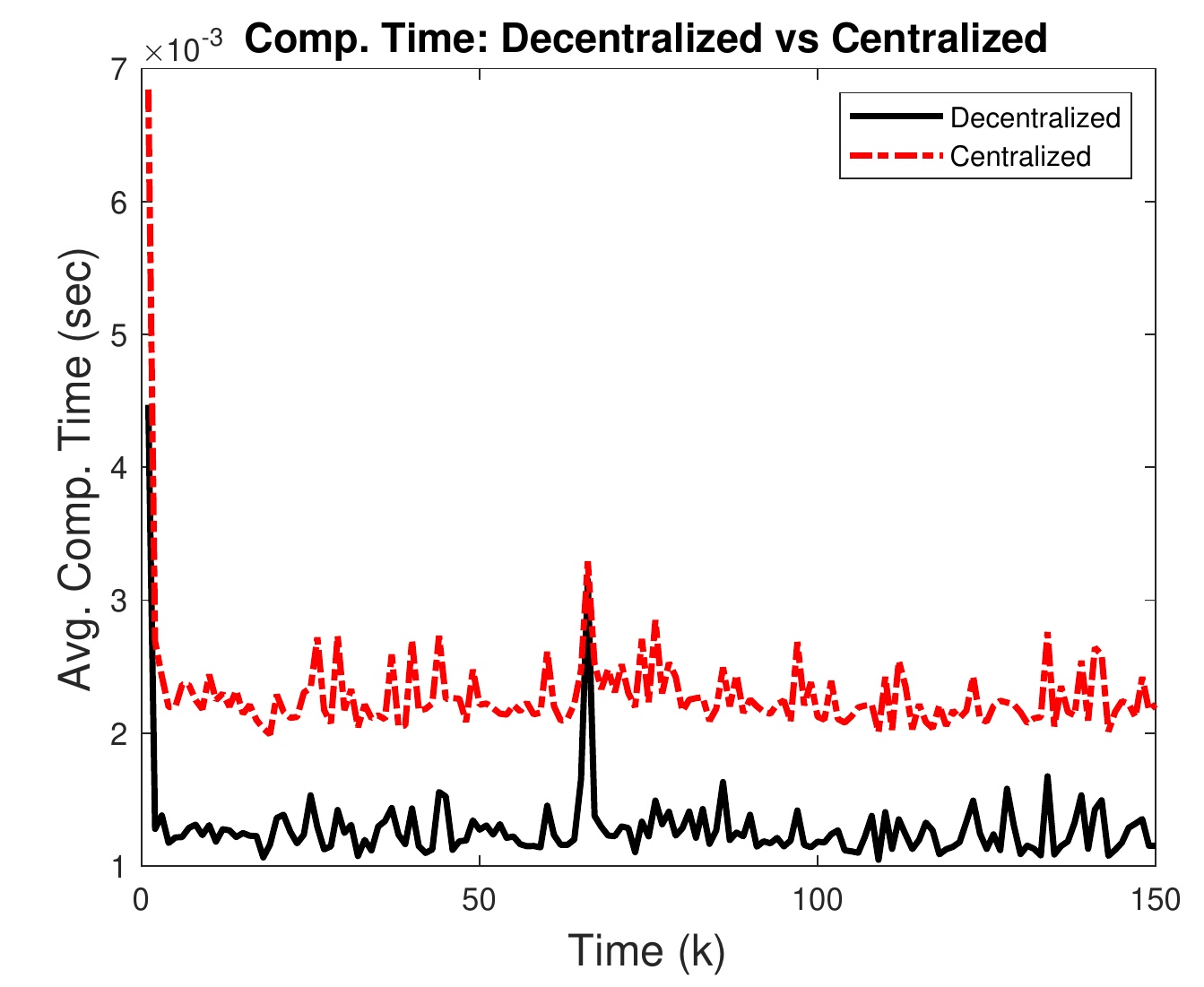}}
\caption{Average computation time: multitarget}
\label{fig: time_multi}
\end{figure}



The tracking performance for a single target tracking and multitarget tracking shows a similar view in the Figure~\ref{fig: error} and \ref{fig: error_multi}. In both cases, the centralized approach outperforms the decentralized approach. Figure~\ref{fig: error} and \ref{fig: error_multi} represent that centralized approach gives less target tracking error for both single and multitarget tracking in all time steps. However, decentralized approach shows contrasting view for single and multitarget tracking. Average target tracking error for multitarget tracking in decentralized setting illustrates increasing trend over discrete time step although the error for single target tracking waves between 15 and 30 meters. For the multitarget tracking in our case, the targets are going away from one another, which leads the UAVs to spread as well. As distance between target and UAVs other than assigned to the target increases over time step, measurement error is expected to be higher which leads the decentralized approach for multitarget tracking error increase over time.

Average computation time for both single and multitarget tracking shows similar view. In both cases, centralized approach takes more time than decentralized approach which is quite expected.



\chapter{Decentralized Data Fusion in UAV Swarm}\label{chapter:consensus}

\section{Introduction}

Autonomous and adaptive sensing has applications such as target tracking, surveillance \cite{surveillance}, autonomous car navigation \cite{auto-navi}, and UAV swarm tactics \cite{ragi-tracking,ragi-dec-pomdp}. Particularly, target tracking via adaptive sensing is becoming increasingly important in autonomous car industry for accurate pedestrian detection and tracking \cite{Blackman}. Sensors such as RADAR, LIDAR, optical sensors, thermal sensors are typically used to measure the target state including its position, velocity, and acceleration. Target tracking with multiple sensors was studied in the past, e.g., \cite{ragi-tracking}, where a central fusion node was responsible for making sensing decisions (e.g., sensor location - assuming sensor mounted on a UAV) for all the sensors combined. Clearly, sensing decisions optimized for all the sensors combined provides the best target tracking performance as these decisions are coupled via sensor data fusion. The main drawback with these centralized decision making methods is that they are computationally intensive as the computational complexity is exponential in the decision space and the number of sensors. To address this challenge, we investigated decentralized strategies in the past to some extent \cite{ragi-dec-pomdp}. 

In this study, we develop a decentralized autonomous sensing method over a networked sensor system for a target tracking application. Specifically, we extend an existing approach called \emph{average consensus algorithm} to perform decentralized data fusion while tracking a moving target. The sensor network is modeled by an undirected graph, which is assumed to be non-time varying. Each sensor generates a noisy measurement of the target state. The presence of an edge between the nodes or sensors means that the sensors are allowed to exchange information/messages for data fusion. In this study, we assume that each sensor maintains a local tracker (or tracking algorithm, e.g., Kalman filter), which updates its local target state estimate using the locally generated sensor measurements and the information it receives from its neighbors. We measure the performance of the above consensus algorithm with \emph{average target tracking error} - the mean-squared error between the target state (ground truth) and the estimate. As a benchmark, we also implement the standard Bayesian data fusion approach for performance comparison. 

The rest of the Chapter is organized as follows. Section~\ref{sec:problem_spec1} presents the problem specification and the objectives. Section~\ref{sec:problem_form1} provides the problem formulation and the methods. We describe simulation results in Section~\ref{sec:sim_result1}. 

\section{Average Consensus Algorithm}
\begin{figure}[H]
\centering{\includegraphics[width= 0.8\columnwidth]{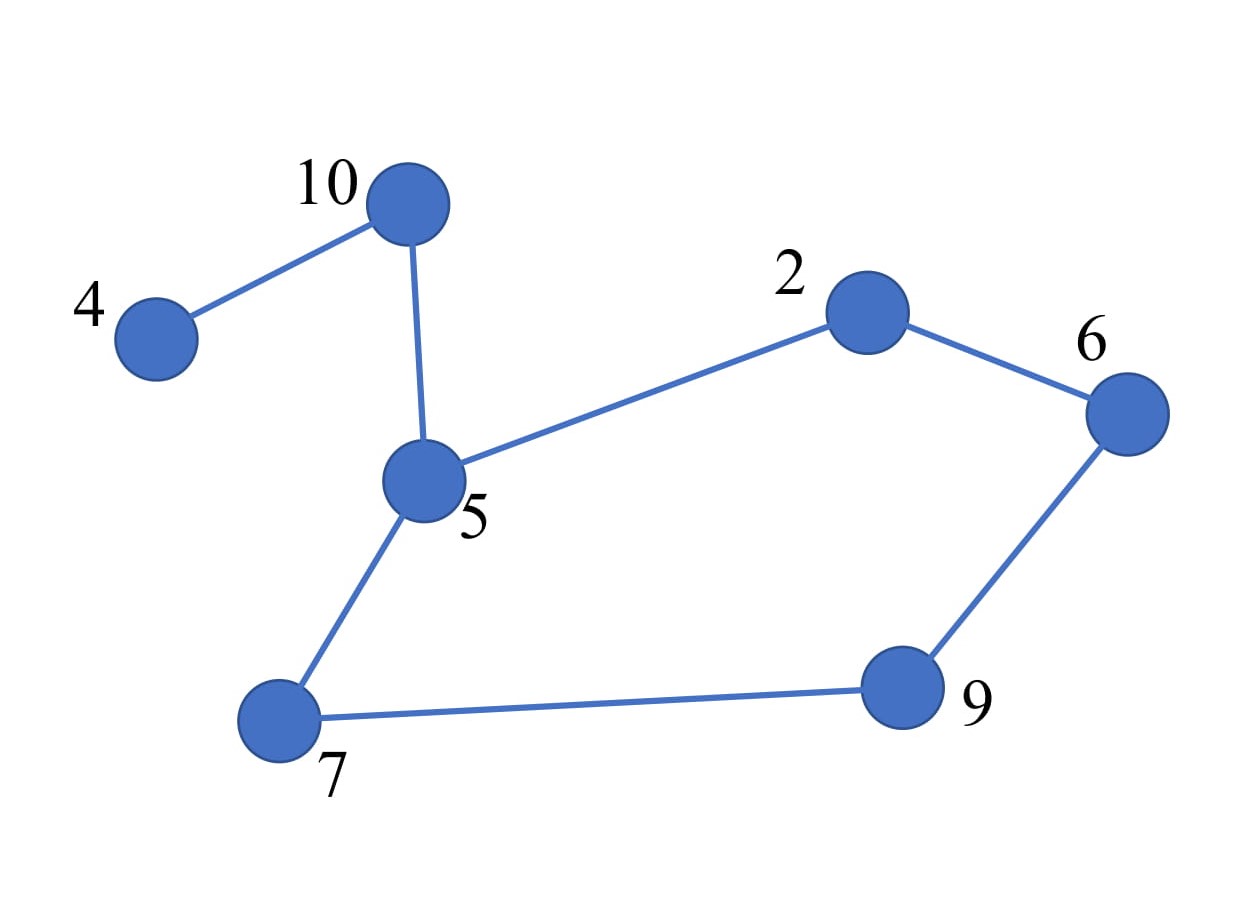}}
\caption{Basic idea of average consensus algorithm - I}
\label{network}
\end{figure}

Assume there are a number of sensors in a sensor network each having it's local attribute (real number). The sensors are connected according to a network graph $G$ at a certain discrete time step $k$. Figure~\ref{network} depicts such a sensor network with different local values at each sensor. Sensors have to come to a consensus on the attribute. 

\begin{figure}[H]
\centering{\includegraphics[width= 0.8\columnwidth]{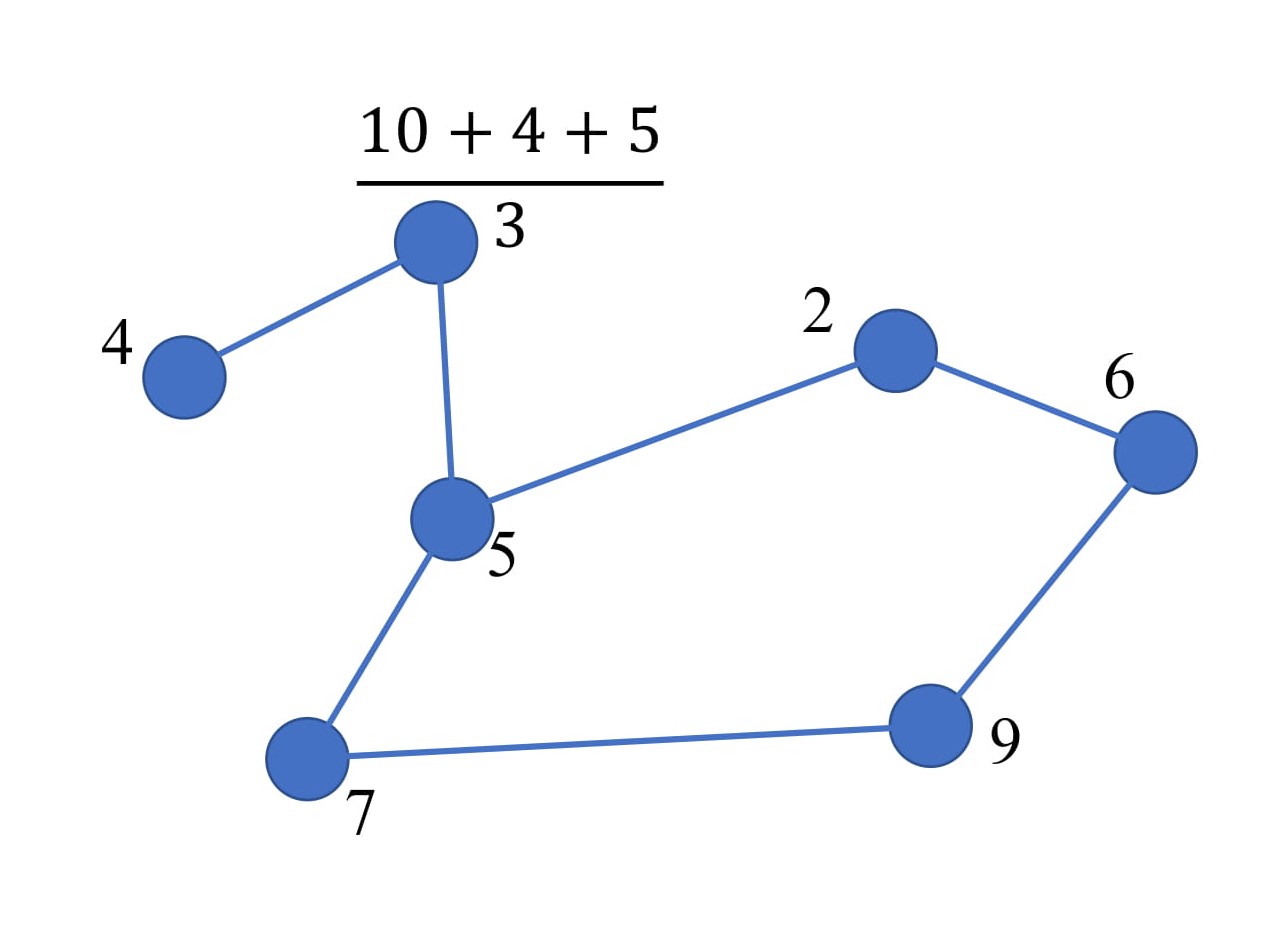}}
\caption{Basic idea of average consensus algorithm - II}
\label{net3}
\end{figure}

If two sensors in the graph $G$ has an edge between them, we call them neighbors. Sensors share their attributes with their neighbors. For a sensor $i$, there are $j$ neighboring sensors communicating to sensor $i$. Each sensor updates its attributes to its neighbors at every discrete time step. Once sensor $i$ has its neighbors attributes at time step $k$, it updates its attribute at time step $k+1$ by taking arithmetic mean of its own attribute and its neighboring sensors attributes. Figure~\ref{net3} shows the sensor with an attribute 10 was updated by $(10+4+5)/3$. 

\begin{figure}[H]
\centering{\includegraphics[width= 0.8\columnwidth]{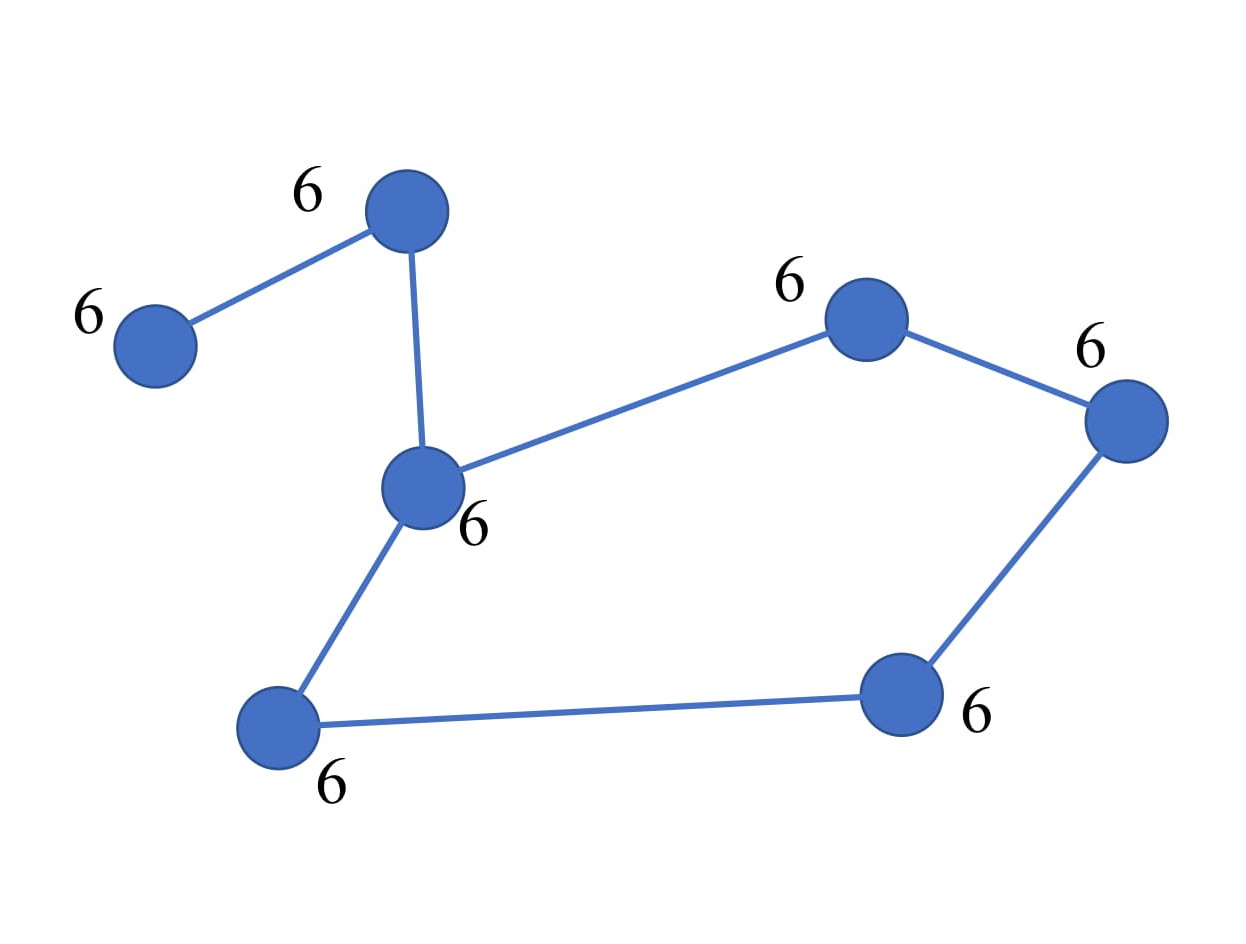}}
\caption{Basic idea of average consensus algorithm - III}
\label{net2}
\end{figure}

The authors of \cite{nedic2010} showed that the sensors updating their attributes reach a consensus after a certain time steps. They also establish a convergence rate estimate. Figure~\ref{net2} represents a network with sensors having a consensus on their attributes. The graph has to be connected for all time steps in order for a consensus to happen.

\section{Problem Specification}\label{sec:problem_spec1}
In our study, we assume there are $n$ sensors tracking a moving target in a decentralized setting, where the sensors are connected via an undirected graph. The target is assumed to be moving on a 2-D plane, where the motion is modeled via a stochastic process, i.e., the state-transition law is a linear model with zero-mean Gaussian noise. We assume the sensor measurement law is also linear with zero-mean Gaussian noise. Thus, each sensor maintains and updates a local target state estimate via Kalman filtering algorithm.

We assume that the sensors have limited battery power and computational capabilities, which sets limitations on the sensors in terms of how they generate measurements and communicate with other sensors. Specifically, we assume that the sensors can either sense (generate target measurements) or exchange information with neighboring sensors, but not simultaneously.

\textbf{Communications:} The sensors have communications capabilities, i.e, each sensor can transmit or receive data to/from the sensors they share edges in the network graph. 
We further assume that the communications delay is negligible.

\textbf{Sensor network:} The $n$ sensors are assumed to be connected via an undirected graph. Each sensor $i$ has a set of neighbors, denoted by $N(i)$, where sensor $j \in N(i)$ if there is an edge connecting $j$ with $i$.

\textbf{Performance measure:} We measure the performance of the algorithms using \emph{average tracking error}, which is the mean-squared error between the target state and the estimates averaged over all the sensors and over time.

\textbf{Objective:} The objective is to compare the performance the \emph{average consensus} algorithm against the standard \emph{decentralized Bayesian data fusion} technique for target tracking with a decentralized sensor network. We measure the performance of these algorithms for different sensor network configurations.

\section{Problem Formulation}\label{sec:problem_form1}

\subsection{Tracking Approach} 

In our study, $\{1,...,n\}$ represent the sensor indices, and $S_i$ represents the 2D location of sensor $i$. The target's motion is described by a linear state-space model (specifically \emph{constant velocity} model \cite{constant-velocity-tracking}):
\begin{equation} \label{equation1}
x_k= Ax_{k-1}+\theta_k, \quad \theta_{k}\sim \mc{N}(0,Q)
\end{equation}
where $x_k$ is the state of the target at time $k$ (which includes the target's 2D location, 2D velocity, and 2D acceleration), $A$ is a state transition matrix, and $\theta_{k}$ is an additive process noise with zero-mean normal distribution with co-variance matrix $Q$. Sensor $i$ generates a position measurement $z_{k}^i$ given by:
\begin{equation} \label{equation2}
z_{k}^i=Hx_{k}+v_{k}^i
\end{equation}
where $H$ is the observation matrix given by 
\[H = \begin{bmatrix}
    1 & 0 & 0 & 0 & 0 & 0\\
    0 & 1 & 0 & 0 & 0 & 0\\
\end{bmatrix},\]
which means that the sensors only generate positional measurements. 

Here $v_{k}^i\sim \mc{N}(0,R(x_k,S_i))$ is the random additive measurement noise modeled as a zero-mean normal distribution, where the co-variance matrix $R(x_k,S_i)$ captures the dependence of the noise characteristics on the location of the target with respect to the sensor. Here, $R_k$ reflects $10\% $ range uncertainty and $0.01\pi$ radian angular uncertainty. Since the state and the observation laws are linear with zero-mean Gaussian noise disturbances, we run Kalman filter at each sensor node to maintain and update the target state posterior distribution with mean and co-variance given by $\hat x_{k|k}^i$ and $P_{k|k}^i$.

Clearly, if the sensors do not exchange any information, the tracking performance suffers at each node. The sensors are connected via an undirected graph, where the presence of an edge between nodes $i$ and $j$ means that the sensors are allowed to exchange information. So, we extend an approach called \emph{average consensus} algorithm to allows sensors to exchange information in a manner that improves the target tracking performance across the sensor network. 

\subsection{Average Consensus} 
Average consensus algorithms let a network of sensors or agents reach a common consensus on certain attributes (real numbers) such as the agent opinions, sensor measurements, etc. Specifically, in these approaches, each agent or sensor updates/replaces (in an iterative manner over time) its local value by taking a weighted average between its local value and the values from all the neighbors. We extend this approach to let the sensors in our problem reach a common consensus on their state estimate parameters (mean vector and covariance matrix). Let $y_{k}^i$ is a vector obtained by concatenating $\hat{x}_{k|k}^i$ and $P_{k|k}^i$ into a column vector at sensor $i$ at time $k$. $N(i)$ is the set of neighbors for $i^{th}$ sensor. Average consensus algorithm applied to our problem is captured by the following equation:
\begin{equation} \label{equation3}
y_{k+1}^i = \frac{ \alpha y_k^i + (1-\alpha)\sum_{j \in N(i)}y_k^j}{\alpha + |N(i)|(1-\alpha)}, \quad \forall  i
\end{equation}
where $\alpha$ is a weighting parameter.

This algorithm achieves its objective if all the sensors reach consensus on the state estimation parameters, i.e., $y_k^i$ = $y_k^j$ for all $i,j$.

\subsection{Decentralized Bayesian data fusion}
Multi-sensor data fusion techniques can be applied in both centralized and decentralized settings. In our study, we use decentralized Bayesian data fusion techniques over the sensor network. Each sensor has a local state estimate $x_k^i$ which is updated in each time step by fusing $x_k^i$ with the estimates from its neighboring sensors as given by the following equations (using standard Bayes rules \cite{fusion-bayesian}).


\begin{equation} \label{equation11}
P_{k+1}^i = \left((P_k^i)^{-1} + \sum_{j = 1}^{N(i)}(P_k^j)^{-1}\right)^{-1}
\end{equation}

\begin{equation} \label{equation12}
\hat{x}_{k+1}^i = P_{k+1}^i\Bigg((P_k^i)^{-1}\hat{x}_k^i +  \sum_{j=1}^{N(i)} ({P_k^i})^{-1} {\hat x_k^j}\Bigg)^{-1}
\end{equation}

\section{Simulation Results}\label{sec:sim_result1}
\begin{figure}
    \centering
    \includegraphics[width=0.55\columnwidth,trim=302 20 210 0, clip]{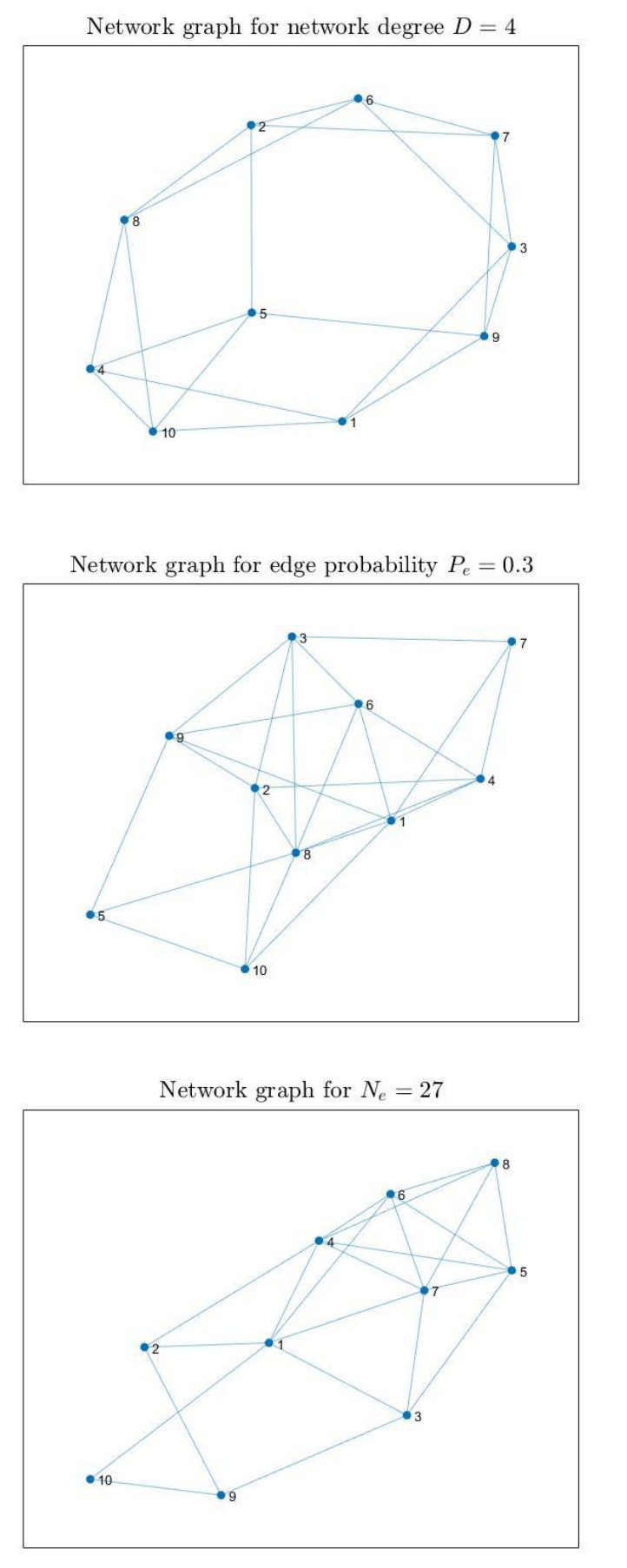}
    \caption{Examples of configurations (configuration I, II, III from top to bottom)}
    \label{fig:my_label}
\end{figure}

We implement our methods for a scenario with 10 sensors, i.e., $n=10$. We set $\alpha = 0.5$ in the following numerical studies except when we evaluate the performance of our algorithms with varying $\alpha$. We compare the performance of the average consensus algorithm against the decentralized Bayesian data fusion approach for different sensor network configurations with \emph{average tracking error} (defined earlier) as the performance measure. In our numerical studies, we use error bars with one standard deviation to show the spread of the performance measure for multiple network graphs generated from a given configuration as discussed below (examples of configurations in Figure~\ref{fig:my_label}).

\textbf{Configuration I}. This corresponds to a network where each sensor has the same degree, where the degree is given by $D$, which is referred to as \emph{network degree}. We generate a random graph with $n$ sensors and $D$ network degree.

\textbf{Configuration II}. In this configuration, we generate a random graph with \emph{edge probability} $P_e$, where $P_e$ represents a probability of an edge existing between two sensors. We start with $n$ sensors with no edges at the beginning, and we create an edge between every pair of sensors with probability $P_e$. We repeat this process until we get a connected network. 

\textbf{Configuration III}. This corresponds to a network with a total number of edges $N_e$ in a connected network.

As sensors typically have limited computational capability and limited battery life, we assume they can run only tracking algorithm while generating sensor measurements or only communicate with neighbors, i.e., run the consensus or data fusion methods as described in Section~\ref{sec:problem_spec1}. Specifically, in our study, sensors track the target for $M$ time steps and apply the consensus/data fusion algorithms in the next $M$ time steps, and repeat the process. During the $M$ time steps when the consensus/data fusion algorithms are being applied, sensors update the state estimates of the target without the measurements, i.e., perform only the prediction step and ignore the measurement update step. In other words, the uncertainty in the target state estimate steadily increases during these $M$ time steps. 

Let $Z$ represent the total number of time steps in our simulation run time. We set $Z=300$ in this study. We define the \emph{average tracking error} measure as follows:
\[\frac{1}{Z}\frac{1}{n}\sum_{k=1}^{Z}\sum_{i=1}^{n} \norm{\hat{x}_k^i - x_k}_2^2 \]
where $x_k$ represents the ground truth at time $k$, and $\norm{\cdot}_2$ is the Euclidean norm.

\subsection{Average tracking error vs.\ $M$}
\begin{figure}[!htbp]
\centering{\includegraphics[width= 0.8\columnwidth, trim = 110 240 120 250,clip]{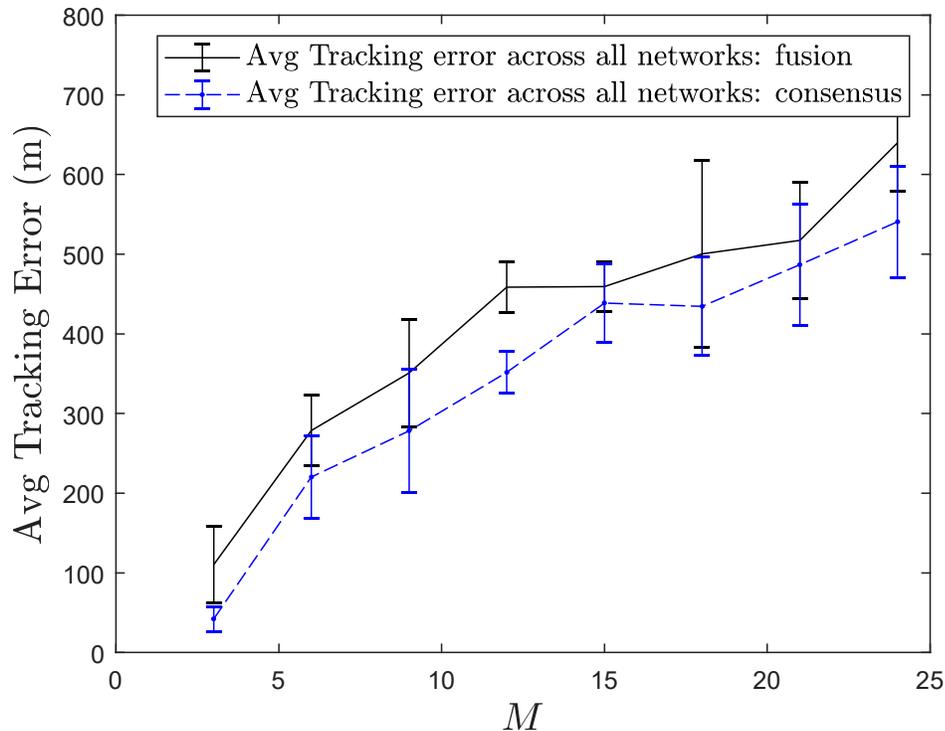}}
\caption{Average tracking error across all sensors with respect to number of time steps, $M$}
\label{fig:Ma}
\end{figure}

We now compare the performance of \emph{average consensus} and \emph{decentralized Bayesian data fusion} algorithms for different values of $M$ on five randomly generated graphs for $n=10$. We evaluate the average tracking error, as defined earlier, for each value of $M$ considered. Figure~\ref{fig:Ma} shows the average tracking error as a function of $M$, where $M\in\{3,6,9,\ldots,24\}$. The figure suggests that the average consensus algorithm outperforms the data fusion approach for all values of $M$ considered. The consensus algorithm seems to be more effective in merging information from multiple sensors than the standard decentralized Bayesian data fusion approach.

\begin{figure}[!htbp]
\centering{\includegraphics[width= 0.8\columnwidth, trim = 110 240 120 250,clip]{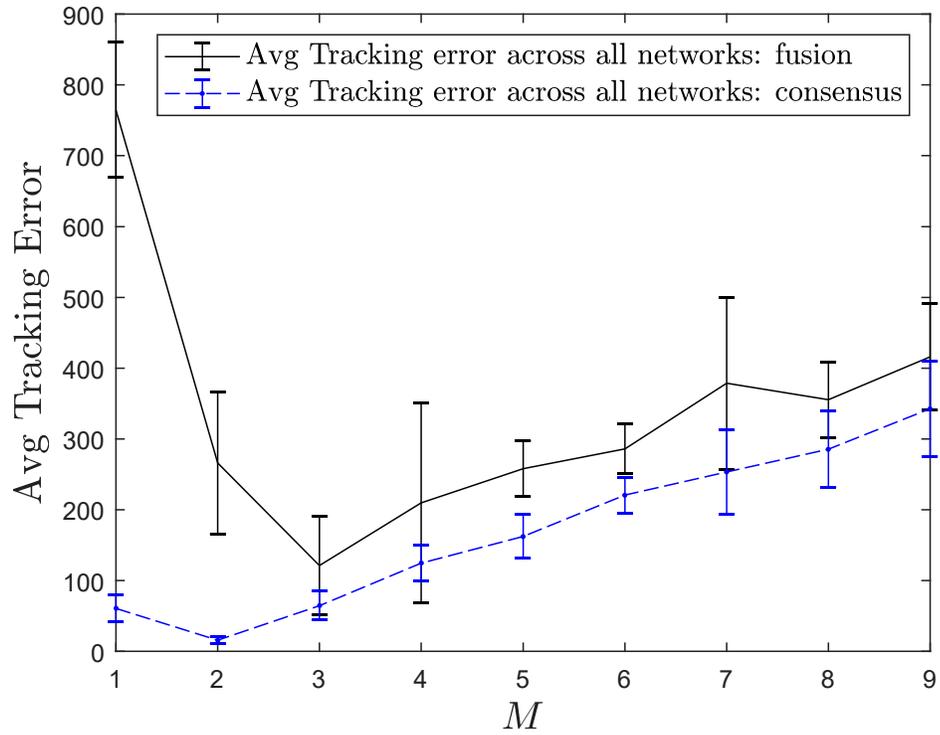}}
\caption{Average tracking error across all sensors with respect to number of time steps, $M$}
\label{fig:Mb}
\end{figure}

Figure~\ref{fig:Mb} represents average tracking error as a function of $M$ for $M\in \{1,2,\ldots,9\}$. Figure~\ref{fig:Mb} shows that the average consensus and decentralized Bayesian data fusion algorithm give better performance for $M = 2$ and $M = 3$ respectively compared to all other values of $M$ considered here.

\subsection{Average tracking error for configuration I}
\begin{figure}[!htbp]
\centering{\includegraphics[width= 0.8\columnwidth, trim = 110 240 120 250,clip]{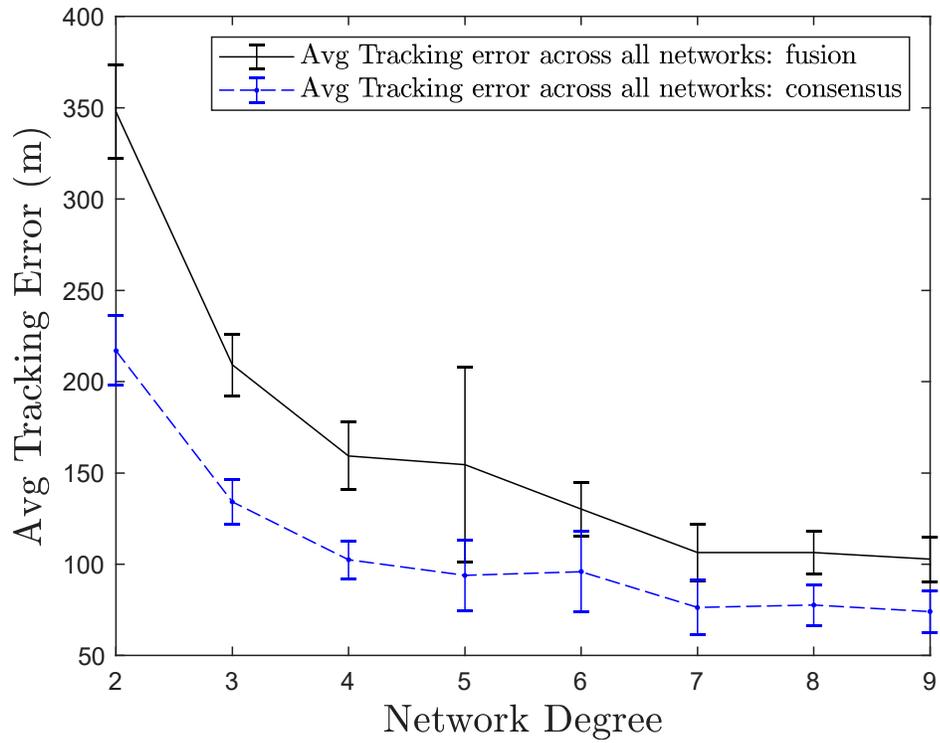}}
\caption{Average tracking error across all sensors for configuration I}
\label{fig:D}
\end{figure}

We now evaluate the average tracking error as a function of the network degree as shown in Figure~\ref{fig:D}. We compare the performance of these two algorithms on five randomly generated graphs for $M=1$ and $n=10$. We observe that the performance of both algorithms increase as the network degree increases. Furthermore, from Figure~\ref{fig:D}, we observe that the average consensus algorithm performs better than the decentralized Bayesian data fusion method. This is an expected behavior since with greater network degree, the sensors have better capability in merging information from other sensors.   


\subsection{Average tracking error for configuration II}
\begin{figure}[!htbp]
\centering{\includegraphics[width= 0.8\columnwidth, trim = 105 240 120 250,clip]{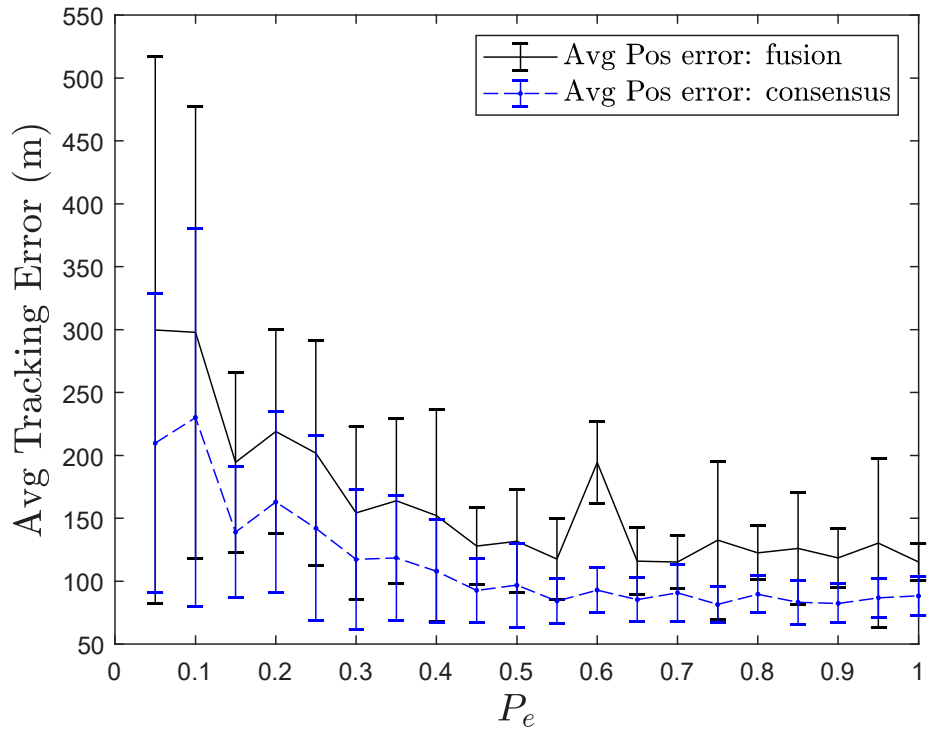}}
\caption{Average tracking error across all sensors with respect to edge probability $P_e$}
\label{fig:P}
\end{figure}

We now perform the same numerical study for a randomly generated graph by using Configuration II with different values of $P_e$ drawn from the set $\{0.1,0.2,\ldots,1\}$. For each $P_e$, we generate 10 graphs. Figure~\ref{fig:P} shows that, for both algorithms, the average tracking error decreases with respect to $P_e$, which is expected since the network connectivity increases with increasing $P_e$. We also notice that the consensus algorithm outperforms the decentralized Bayesian data fusion approach for each $P_e$.  

\subsection{Average tracking error for configuration III}
\begin{figure}[!htbp]
\centering{\includegraphics[width= 0.8\columnwidth, trim = 110 240 120 250,clip]{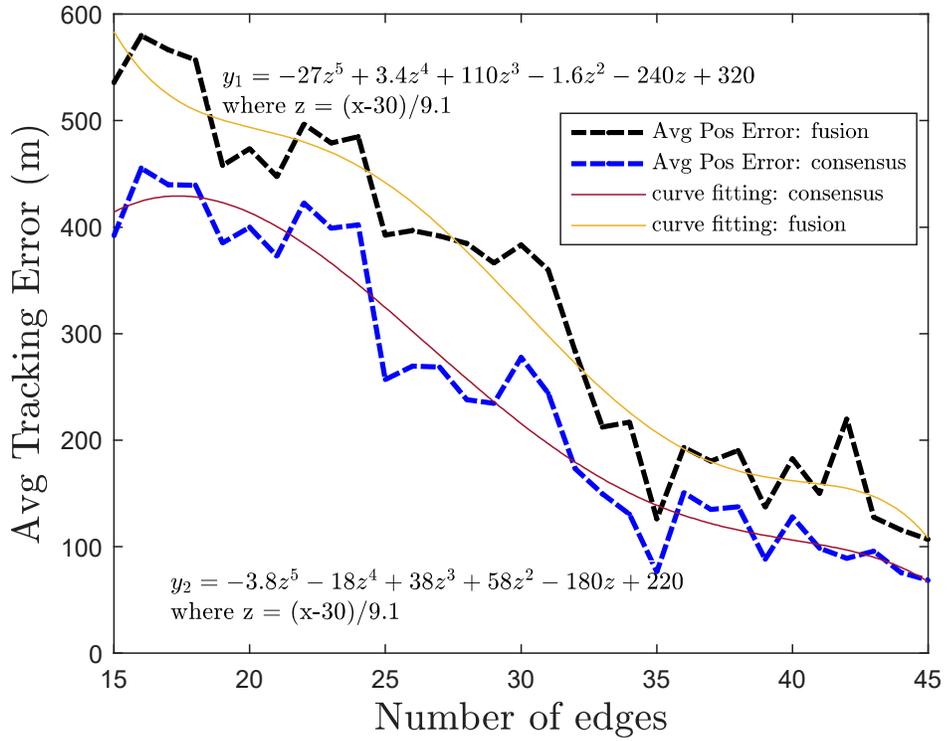}}
\caption{Average tracking error across all sensors with respect to number of edges}
\label{fig:Ne}
\end{figure}

We now evaluate the average tracking error for different value of $N_e$ as shown in Figure~\ref{fig:Ne}. We generate (randomly) five graphs with Configuration~III for this study. We observe that with increasing $N_e$, the performance of both of the algorithms increases. We fit $5^{th}$ degree polynomial curves for the performance plots in Figure~\ref{fig:Ne}, which characterize the variation of the performance of the algorithms as a function of $N_e$.  

\subsection{Average tracking error for weighting parameter $\alpha$}

\begin{figure}
\centering{\includegraphics[width= 0.8\columnwidth, trim = 110 240 120 250,clip]{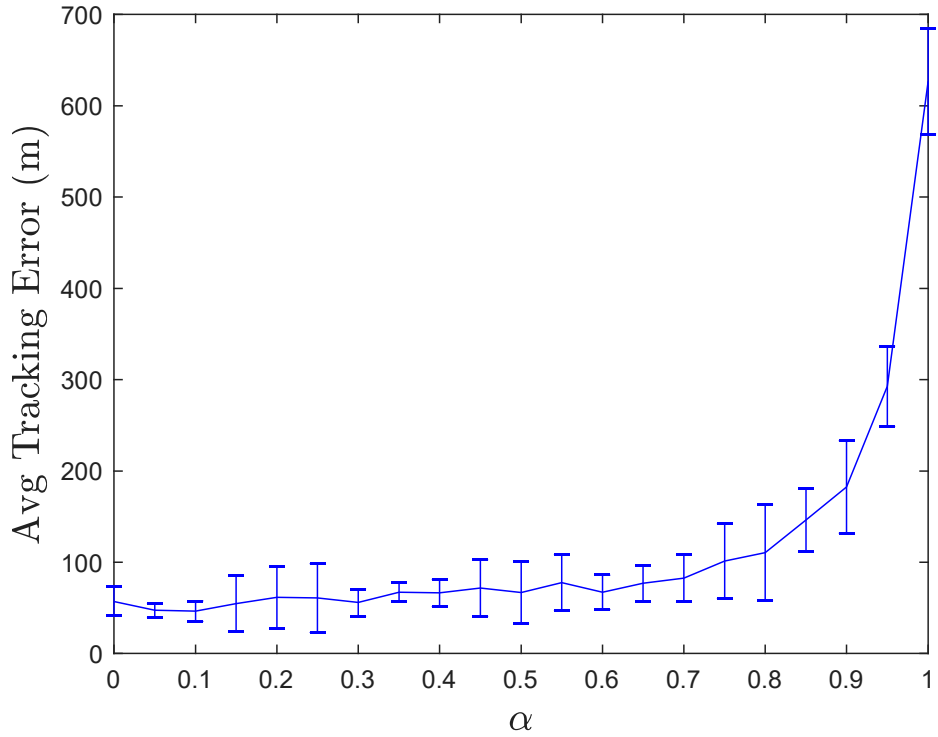}}
\caption{Average tracking error across all sensors with respect to weighting parameter $\alpha$}
\label{fig:alpha}
\end{figure}

In this part, we study the performance of the average consensus algorithm with respect to the weighting parameter $\alpha$. Here, $\alpha = 0$ means that the consensus algorithm replaces the local sensor's state estimate with the average of its neighbors' estimates. On the other hand, $\alpha = 1$ means that the consensus algorithm ignores the estimates from the neighbors and simply retains the local state estimate. For different values of $\alpha$ in the interval $[0,1]$, we evaluate the average tracking error, as shown in Figure~\ref{fig:alpha}. The figure shows that the average tracking error increases significantly when the value of $\alpha$ is close to $1$.




\chapter{Conclusions}

In this thesis, we developed decentralized control and information fusion methods for UAVs in the context of formation control and target tracking applications. Specifically, we extended a decision-theoretic formulation called \emph{decentralized Markov decision process} (Dec-MDP) to develop near real-time decentralized control methods to: a) drive a UAV swarm from an initial formation to a desired formation in the shortest time possible; b) drive a UAV swarm to track multiple moving targets while maximizing the tracking performance and avoiding collisions. 

As decision-theoretic approaches suffer from the curse of dimensionality, for computational tractability, we extended an approximate dynamic programming method called \emph{nominal belief-state optimization} (NBO) to solve the Dec-MDP approximately. For benchmarking, we also implemented a centralized approach (Markov decision process-based) and compared the performance of our decentralized control methods against the centralized methods. In the context of the formation control problem, our results show that the average computation time for obtaining the optimal controls and the time taken for the swarm to arrive at the formation shape are significantly less with our Dec-MDP approach compared with that of the centralized methods. Similarly, in the context of the target tracking problem, our Dec-MDP methods significantly outperformed the centralized methods in terms of the computational time required to obtain the optimal controls, while sacrificing only marginally on the target tracking performance.

To address the problem of information fusion in decentralized swarm systems, we extended the \emph{average consensus} algorithm for decentralized data fusion over a networked sensor system for a target tracking application. We compared the performance of our average consensus-based data fusion method against the standard Kalman filter-based centralized data fusion for different network configurations. We found that the average consensus algorithm outperformed the Kalman filter-based data fusion in terms of the target tracking performance.

\section{Future Scope}
The formation control approach discussed in this thesis can be extended to 3D formation, and these formations can be used to sense the environments for 3D reconstruction of a scene. The vantage points of the UAVs in the swarm in 3D formation can be exploited for efficient reconstruction of the scene in 3D, while extending tomography-type approaches. The decentralized control strategies presented in this thesis can be extended to control the motion of the UAVs in the swarm to maximize the efficiency of the above 3D scene reconstruction process. These methods have several applications including the use of drones to map unexplored and unsafe regions (e.g., caves, underground mines, toxic environments).  

In this thesis, we studied decentralized data fusion methods over a time-invariant sensor network. Such methods are critical for inducing cooperative behavior among the agents in a swarm. These decentralized fusion methods can be extended to more realistic time-varying networks, where the future behavior of the network may need to be incorporated in the motion planning of the agents in the swarm for long-term data fusion performance.   

For both the case studies we considered in this thesis: decentralized UAV motion control for formation and multitarget tracking, which are posed as Markov decision processes, are solved via an approximate dynamic programming approach NBO. At the expense of increased computational intensity, other ADP approaches can be extended such as policy rollout, Q-learning, and Monte-Carlo tree search, to improve the optimality of the control decisions.

\supplementaries

\bibliographystyle{IEEEtran}
\bibliography{harvey,Subspace2017}

\begin{vita}

Md Ali Azam was born in Pabn, Bangladesh. He earned his Bachelor of Science in Electronics and Telecommunication Engineering from Rajshahi University of Engineering and Technology (RUET), Bangladesh. He completed his Master of Science in Electrical Engineering from South Dakota School of Mines and Technology (SDSMT). 

He was a graduate assistant at SDSMT where he worked as a graduate teaching assistant and graduate research assistant during his MS studies. He also worked as a safety officer at Public Safety Department at SDSMT. He worked as a system engineer at a telecommunication company in Bangladesh before attending SDSMT. 

During his MS studies, he published two conference papers as a leading author and one conference paper as a co-author. He is a student member of IEEE and SDSMT Graduate Student Society.

\end{vita}

\end{document}